\documentclass[11pt,a4paper]{article}

\usepackage[utf8]{inputenc}
\usepackage[T1]{fontenc}
\usepackage{textgreek}

\usepackage{amsmath}
\usepackage{amssymb}
\usepackage{amsfonts}
\usepackage{bm}

\usepackage{graphicx}
\usepackage{svg}
\usepackage{float}
\usepackage{subcaption}
\usepackage{svg}
\usepackage{booktabs}
\usepackage{multirow}
\usepackage{array}

\usepackage{siunitx}

\usepackage{xcolor}
\usepackage[nopatch=footnote]{microtype}

\usepackage[
    colorlinks=true,
    linkcolor=blue,
    citecolor=blue,
    urlcolor=blue
]{hyperref}

\usepackage{cleveref}

\usepackage{graphicx}
\usepackage{amsmath}
\usepackage{amssymb}
\usepackage{booktabs}
\usepackage{caption}
\usepackage{float}
\usepackage{siunitx}
\usepackage{setspace}
\usepackage{hyperref}

\usepackage{tensor}
\usepackage{bm}
\usepackage{physics}
\AtBeginDocument{\RenewCommandCopy\qty\SI}
\ExplSyntaxOn
\msg_redirect_name:nnn { siunitx } { physics-pkg } { none }
\ExplSyntaxOff

\newcommand{\Heff}{H_{\mathrm{eff}}}
\newcommand{\HTC}{H_{\mathrm{TC}}}
\newcommand{\Hcav}{H_{\mathrm{cav}}}
\newcommand{\Hatm}{H_{\mathrm{at}}}
\newcommand{\Dc}{\Delta_c}
\newcommand{\Da}{\Delta_a}
\newcommand{\Tp}[1]{\hat{T}_{#1}^{+}}
\newcommand{\Tm}[1]{\hat{T}_{#1}^{-}}
\newcommand{\sip}[1]{\hat{\sigma}_{+}^{(#1)}}
\newcommand{\sm}[1]{\hat{\sigma}_{-}^{(#1)}}
\newcommand{\siz}[1]{\hat{\sigma}_{z}^{(#1)}}
\newcommand{\Dket}[1]{\ket{D_{#1}}}
\newcommand{\Bket}[1]{\ket{B_{#1}}}
\newcommand{\adag}{\hat{a}^{\dagger}}
\newcommand{\ah}{\hat{a}}
\newcommand{\Rop}{\hat{R}}
\newcommand{\Proj}{P}
\newcommand{\Qroj}{Q}
\newcommand{\Id}{\hat{\mathbf{1}}}
\newcommand{\tx}{\hat{\tau}_x}
\newcommand{\ty}{\hat{\tau}_y}

\newcommand{\kpar}{k_{\parallel}}
\newcommand{\kp}{k_{+}}
\newcommand{\km}{k_{-}}
\newcommand{\kpm}{k_{\pm}}
\newcommand{\wt}{\omega_3}
\newcommand{\vtri}{v_{\triangle}}

\newcommand{\kem}{\bm{k}_{\rm em}}

\newcommand{\vt}{v_\triangle}
\newcommand{\om}{\omega}
\newcommand{\omo}{\omega_0}
\newcommand{\phiL}{\phi_{k_L}}
\newcommand{\phib}{\phi_b}
\newcommand{\psiket}{|\psi_0\rangle}
\newcommand{\Epm}{E_\pm}

\newcommand{\no}{n_o}
\newcommand{\hbne}{n_e}
\newcommand{\Dz}{\Delta_0}
\newcommand{\Itot}{I_{\rm tot}}

\usepackage[
    margin=0.9in
]{geometry}

\usepackage[numbers,sort&compress]{natbib}

\begin{document}

\begin{center}

    {\Large \textbf{Extraordinary Optical Spin Hall Effect in 2D Exciton Emission in Open Cavities  }}\\[0.3cm]

    {\large
    Kunal Banerjee$^{1}$, Ben Johns$^{1}$, Nitin Yadav$^{1}$, Anil Shaji$^{*2,3}$, Jino George$^{*1}$
    }\\[0.2cm]

    {\normalsize
    $^{1}$\textit{Molecular Strong Coupling Lab, Department of Chemical Sciences, Indian Institute of Science Education and Research (IISER) Mohali, Punjab, India - 140306.}\\
    $^{2}$\textit{School of Physics, Indian Institute of Science Education and Research (IISER) Thiruvananthapuram, Vithura, Kerala, India – 695551.}\\
    $^{3}$\textit{Centre for High Performance Computing, Indian Institute of Science Education and Research (IISER) Thiruvananthapuram, Vithura, Kerala, India - 695551.}
    
    }

\end{center}

\begin{center}

\begin{abstract}

Open cavities are emerging as a promising platform for studying light-matter interactions. A lossy open cavity couples to an atomic monolayer, resulting in parabolic dispersive emission at room temperature. Here, light-matter interaction is weak, and the dispersion arises from a phenomenon known as dark strong coupling. The bright 2D exciton rapidly decays through the cavity mode, leaving behind spin-dark excitons of the 2D lattice, thereby undergoing symmetry breaking. Further, mapping the polarisation dependence in the k-space reveals spin-orbit interaction in the coupled system. A synthetic spin-orbit Hamiltonian is constructed to extract the effective contributions of Rashba- and Dresselhaus-type interactions in the open cavity. The system shows a clear evolution of the spin-orbit effect at room temperature. Here, emission is polarisation-sensitive, yielding high degrees of linear and circular polarisation. This is one of the simplest configurations for studying photonic spin-orbit coupling, in which an atomic monolayer provides Rashba and Dresselhaus coupling at room temperature. Therefore, an open cavity platform is promising for polarisation-sensitive quantum measurements and device applications. 
\end{abstract}
\end{center}



\section{Introduction}

Light-matter interaction in 2D materials is unique and has recently been applied in various platforms\cite{Ebbesen2016, Lidzey1998}. 2D materials like transition metal dichalcogenides (TMDs) are the best in this category due to their high oscillator strength\cite{Johns2024, Dufferwiel2015, Liu2015}. There is a surge in fundamental studies of polariton physics, including applications in lasing, condensation, photodiodes, photodetectors, and long-range transport using TMDs\cite{Eizner2018,Bhatt2023, Kavokin2022, Zasedatelev2019, Deng2010, Demirchyan2014}. Very recently, open cavities have become more attractive due to their fabrication simplicity and ease of optical measurement\cite{Yadav2025}. Various attempts have been made to prepare one-mirror or mirrorless cavities for polariton studies\cite{Kaur2023}. Nanoparticle-on-mirror geometry is also explored to study molecules and materials at the single-molecule level. Plexitons are another category and have been found to be promising for controlling chemical and physical properties\cite{Li2023}. However, these systems exhibit inhomogeneously broadened absorption line shapes, despite their interesting properties. Recently, we proposed a high-refractive-index layer on top of a mirror to form an open cavity that confines the electromagnetic field at the interface. Here, the system operates in the dark strong coupling (DSC) regime, in which the photoluminescence emits through the cavity channel\cite{Liu2021}. A thick hBN layer on a perfect mirror confines cavity modes with Q-factor in the range of 20-30. Here, $WS_2$ monolayer is used as an active layer for light-matter coupling. Cavity mode can be precisely tuned by adjusting the hBN thickness. It constitutes a non-Hermitian lossy system that interacts directly with the environment. 
Recent directions in light-matter interaction include chiral cavities, chiral discrimination, and optical spin control\cite{Mauro2023,Mauro2024,Bliokh2015}. One such process is the optical spin Hall effect (OSHE)\cite{Leyder2007, Ren2025,Rechcinska2019,Tercas2014,Liu2017,Hosten2008,Spencer2021}. This is analogous to an electron moving in a semiconductor with a specific spin selection rule. For a photon, opposite spin angular momenta correspond to the two types of circular polarisations of light: left (LCP) and right (RCP). Similar to the spin Hall effect of electrons\cite{Dyakonov1971}, when the photon experiences a spin (polarisation) dependent lateral deflection induced by an effective spin-orbit interaction\cite{Sala2015}, it leads to OSHE. The OSHE can be used in various photonics applications, including polarisation-dependent transport and polarised photodetectors. The effect can be engineered by various mechanisms of spin-orbit coupling, such as splitting into transverse electric (TE) and transverse magnetic (TM) modes-Dresselhaus coupling\cite{Hu2003}, and/or Rashba coupling\cite{Ni2023,Kedziora2026,Li2022,LiangQ2024,LiangJ2024, Sedov2024, Shelykh2018, Ling2015, Ren2021, Liang2026} in an anisotropic crystal. OSHE has been observed in many configurations, including high-finesse DBR cavities with liquid crystals, perovskites, and organic crystals\cite{LiangQ2024, Spencer2021, Rechcinska2019}.
In this article, we demonstrate OSHE in open optical cavities through a dispersive emission. Here, the bright state decays rapidly leaving behind the spin-dark exciton manifold that emits via the cavity channel in a two-step process\cite{Liu2021}. Interestingly, the emission features are highly polarisation selective, especially in higher k-space. Both linear and circular polarisation studies are conducted and compared as the mode position is varied. A detailed quantum mechanical calculation was used to extract the Rashba and Dresselhaus coupling terms. This provides a clear understanding of how a synthetic Hamiltonian can be used to predict the spin-orbit coupling in an atomic monolayer-open cavity system. Further experimental studies suggest that the Rashba coupling constant is approximately 24 eV Å.   Our theoretical calculations and k-space tomography measurements indicate that a Rashba-type spin texture from the $WS_2$ drives the Dresselhaus interaction arising from the intrinsic TE-TM splitting of the hBN layer, generating a four-lob mixing. Here, we used a simple open cavity configuration to study OSHE in an atomic monolayer at room temperature.

\section{Results and Discussion:}
\begin{figure}[h]
    \centering
    \includegraphics[
        width=0.95\linewidth
    ]{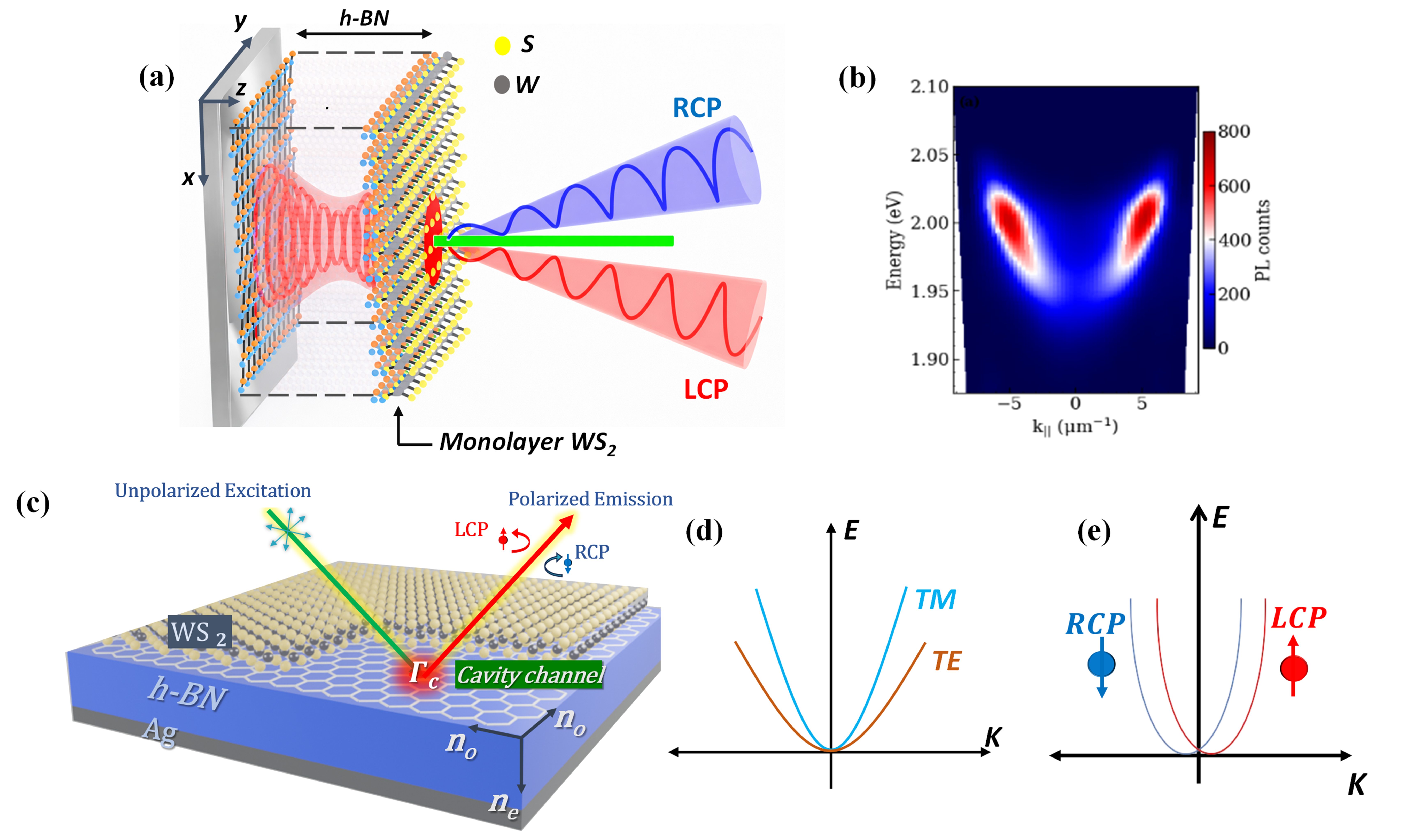}

    \caption{
        Optical spin-Hall effect in polariton-like emission: (a) Schematic of the open cavity heterostructure with $WS_2$ monolayer. The open cavity is made of a silver mirror (120 nm) at the bottom with hBN/air interface on top. (b) Typical polariton-like emission from this structure. (c) Schematic of the chiral induction at the hBN/$WS_2$ interface and a prominent active emission from the cavity channel. (d) The dispersion of the TE-TM components of light reveals their non-equivalence in the k-space. (e) Spin-polarised dispersion band with opposite transverse shift in the k-space.
    }

\end{figure}

In the open cavity structure, the $WS_2$ monolayer is placed at the very top of the heterostructure, whereas the actual cavity is formed between the silver mirror and hBN/air interface. The high refractive index of the hBN layer creates a strong index mismatch with air, which is crucial for confining the electromagnetic mode. This simple configuration allows us to couple the 2D exciton directly to the region of maximum electric field (Figure 1a). In this configuration, at a certain thickness of hBN, the system crosses an exceptional point (EP), and the polariton branches evolve. The EP estimated for such a configuration is approximately 1100 nm of hBN thickness\cite{Johns2026}. Above the EP, the emission shows a parabolic dispersion as shown in Figure 1b. Interestingly, spectral Rabi splitting is absent, indicating that the system enters the so-called DSC regime.  In the DSC regime, the emission can exit the cavity through two possible pathways: the cavity channel or the exciton channel\cite{Liu2021}. The cavity channel overtakes all other emission processes, and we observe a clear parabolic dispersion that can be mapped in k-space.

\begin{figure}[h]
    \centering
    \includegraphics[
        width=0.95\linewidth
    ]{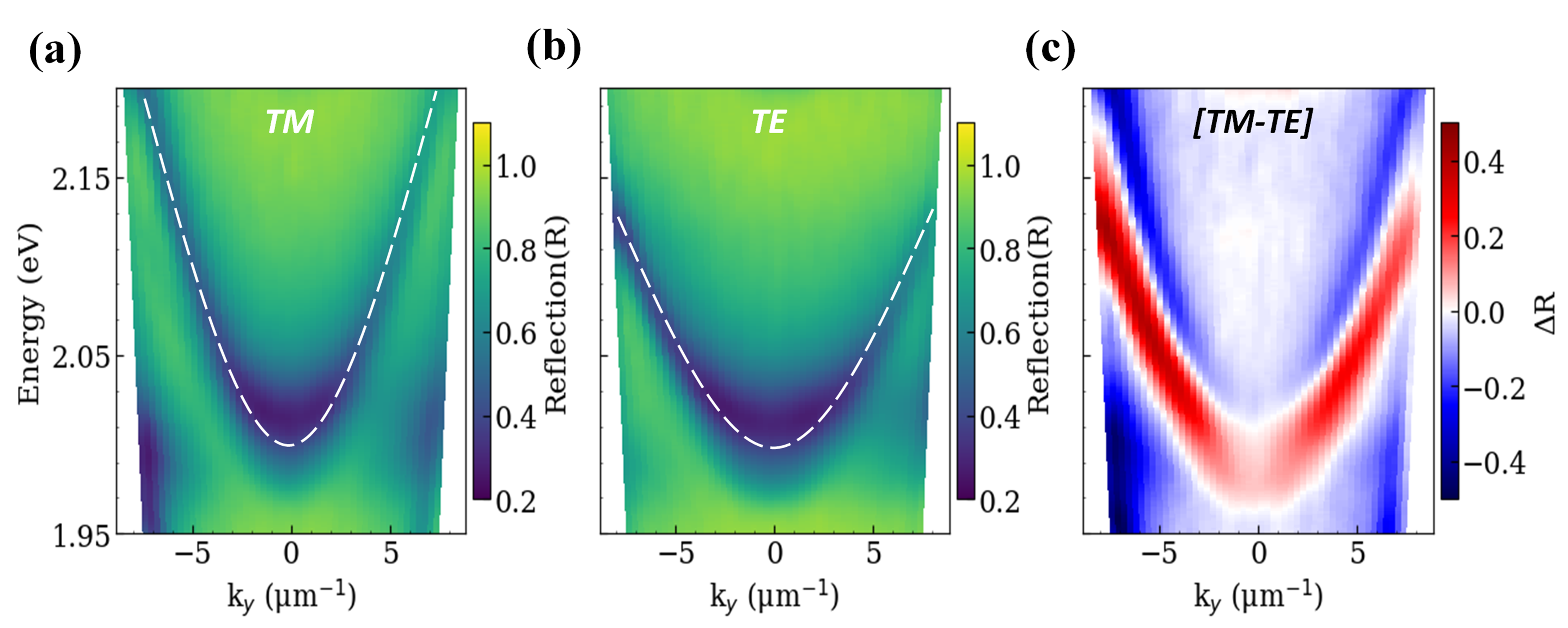}

    \caption{
        TM-TE imbalance in open cavity: a) TM and b) TE component of the experimental reflection spectra collected for the Ag-hBN-$WS_2$ (multilayer) heterostructure, and (c) the corresponding TM-TE difference plot for better comparison.
    }

\end{figure}

The Dresselhaus effect is one of the governing spin-orbit interactions in this open cavity structure, which is prominent due to the TM-TE mode splitting. Polarisation-resolved reflection measurements have shown a clear signature of the photonic spin-orbit interaction in the Ag/hBN/$WS_2$-based open cavity (Figure 2). This effect arises from the intrinsic properties of hBN, specifically due to its out-of-plane anisotropy\cite{Segura2018}. In the reflection measurement of exfoliated bare hBN, the TM and TE components exhibit different parabolic trajectories. Thus, it generates an effective spin-orbit coupling. The estimated TM-TE splitting reaches up to 50 meV at higher k-space, with splitting strength $\beta=10^5$ eV \AA$^2$ . An hBN thickness of 2240 nm has been used in this case. Whenever the cavity wavelength crosses the $WS_2$ absorption wavelength, a much stronger absorption is observed due to the increased density of states (Purcell effect). This effect is stronger if a multilayer is introduced, as shown in the present case. The TM component of the cavity mode grows more rapidly than the TE component, as in the bare hBN cavity mode. As a result, stronger satellite absorption at 2.01 eV is observed at higher momenta for the TM mode, unlike for the TE mode. 
\\
In the next step, circular polarisation measurements were conducted to understand the effect of spin-orbit interaction on the analyser frame. Here, LCP and RCP components are well resolved, with a clear OSHE signature in the system (Figure 3). It has been found that, in the circular polarization resolved reflection measurement, the spin-polarised dispersion bands of opposite helicity do not overlap with each other at $k_x$= 0. The two bands corresponding to the LCP and RCP components shift in opposite momentum directions (along $k_y$). The same measurement has also been performed on the emission k-space, and the band shift occurs in the same way as in the reflection data (Figure 3d-f). In both cases, the difference spectra show an asymmetric distribution in the $k_y$-space, which solidifies our claim of opposite shift in the band. 

\begin{figure}[t]
    \centering
    \includegraphics[
        width=0.9\linewidth
    ]{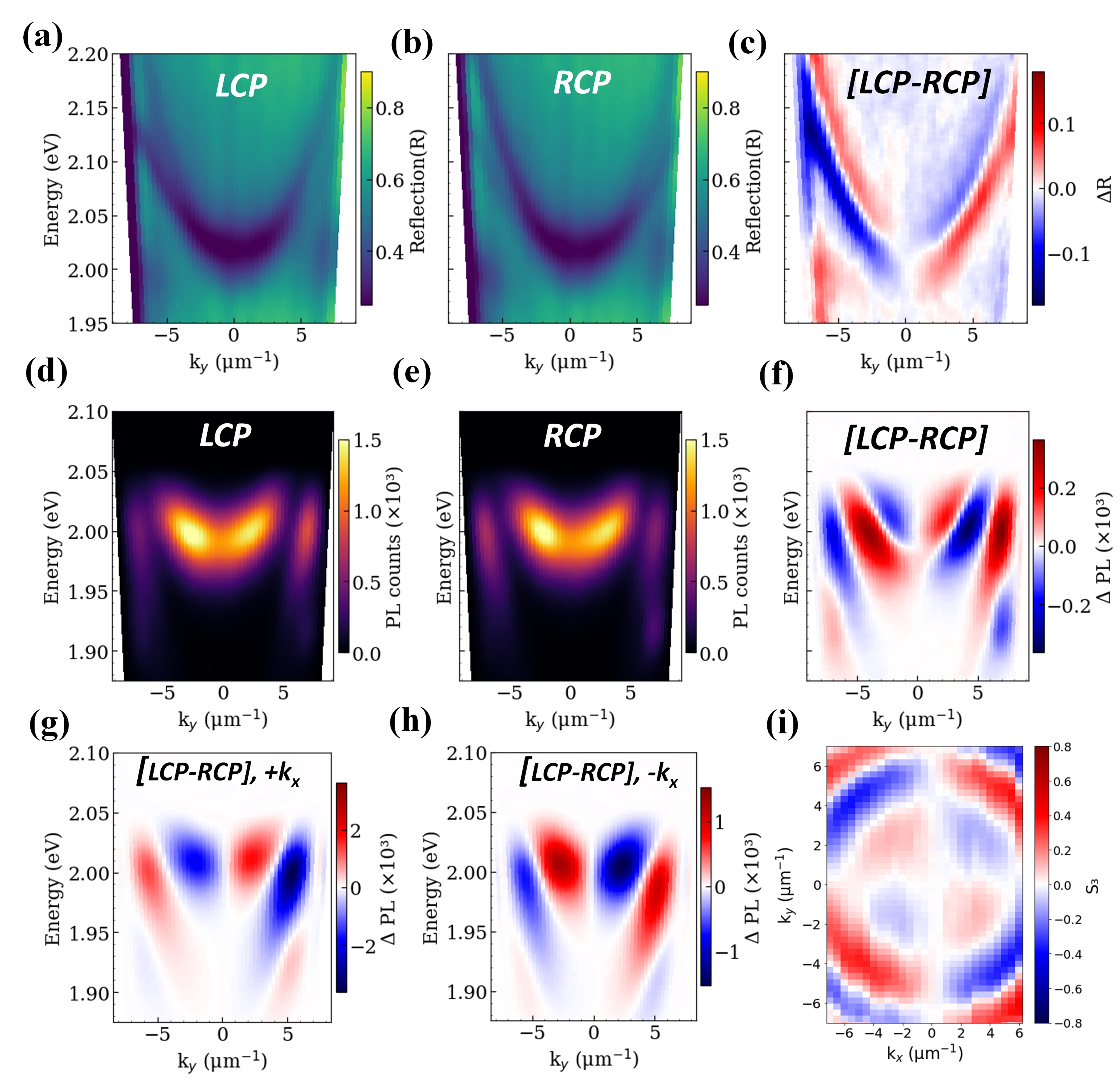}

    \caption{
        Circular polarisation mapping in the k-space: Reflection spectra of (a) LCP and (b) RCP components collected from the open cavity heterostructure and the corresponding difference spectra of (c) LCP-RCP. Polariton-like emission spectra of (d) LCP and (e) RCP components, and the corresponding (f) difference spectra (LCP-RCP) at $k_x$=0. Mapping the $k_x$-space at (g) $k_x$= + 4 $\mu m^{-1}$ and (h) $k_x$= - 4 $\mu m^{-1}$, and the (i) normalised S3-Stokes parameter plotted at 2.01 eV for $k_x$ versus $k_y$ dispersion diagram.
    }

\end{figure}

Further, we took a cross-section at 2.01 eV for both dispersion spectra to visualise the relative shift. From the cross-section plot, we estimated a shift of about 0.619 $\mu m^{-1}$ between the bands.  Besides the $k_x$=0 case, we also tried to capture the momentum space 3D-tomographic data by a transverse shift of the back focal lens with respect to the spectrometer slit. The difference spectra for two opposite higher $k_x$ values (± 4 $\mu m^{-1}$) are shown in Figures 3g and 3h. These two spectra show completely opposite trends when compared. Additionally, the $k_x$ versus $k_y$ spectra of the S3 Stokes polarisation component show a clear four-lobed geometry, a dominant signature of OSHE (Figure 3i).

We have experimentally demonstrated that an open system can generate linear and circularly polarised light at higher momenta. TM-TE components of the dispersive emission for two different hBN thicknesses (1460 nm and 2240 nm) are obtained at $k_x$=0 (Figure 4). The TM component of the cavity mode disperses quickly and reaches the exciton energy within accessible in-plane momenta, resulting in emission from the cavity channel in the DSC regime. 

\begin{figure}[t]
    \centering
    \includegraphics[
        width=0.95\linewidth
    ]{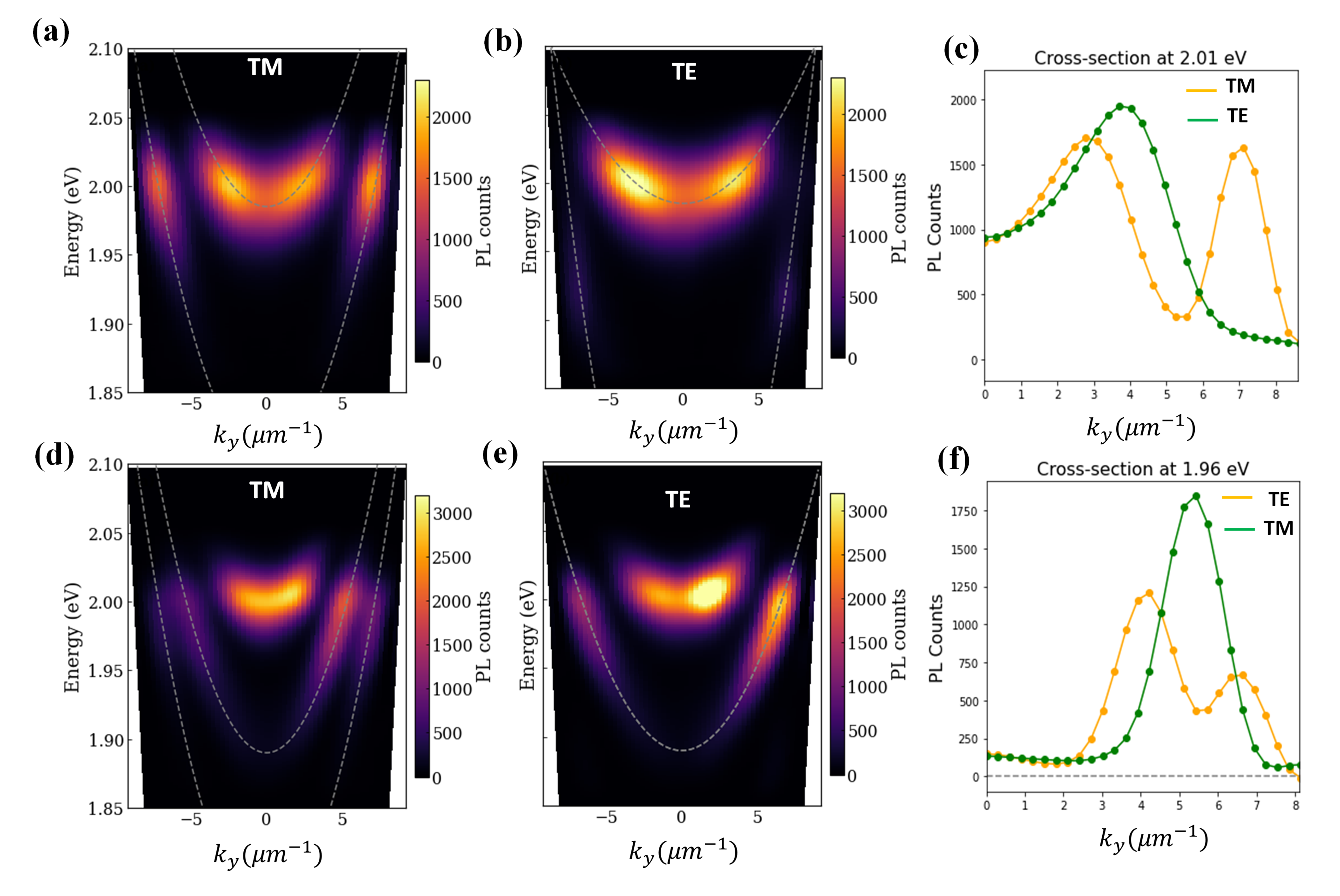}

    \caption{
        TM-TE polarised polariton-like emission: (a, b) TM-TE component of the k-space emission of the heterostructure with hBN thickness 1460 nm and the corresponding (c) cross-section plot at the fixed energy of 2.01 eV. (d, e) A similar plot for the heterostructure with hBN thickness 2240 nm, and the corresponding (f) cross-section plot at 1.96 eV for better comparison of the higher k-space.
    }

\end{figure}

For better clarity, we have taken a momentum cross-section at a suitable energy to compare the temporal evolution of the spectra. We can clearly see that emission from the different modes is linearly polarised. For example, the emission from the heterostructure for 1460 nm hBN thickness at momenta of 6.95 $\mu m^{-1}$ (2.01 eV) has a degree of polarisation (DOP) of 74 . Here, the DOP is calculated as, DOP = $\frac{I^{TE}-I^{TM}}{I^{TE}+I^{TM}}$ at specific momentum and energy values. To achieve circularly polarised luminescence (CPL) with a high extinction ratio, we need to map at a higher $k_x$ space. A clear visualisation of this effect is given in the Appendices. The CPL helicity of the system is interchanged with respect to the sign of $k_x$ and the selection component (LCP or RCP). The extinction value of the CPL generation of one of such bands is approximately 72, which is also calculated as the degree of circular polarization, DOC = $\frac{I^{LCP}-I^{RCP}}{I^{LCP}+I^{RCP}}$. Thus, the heterostructure can produce both linear and circularly polarised light from unpolarised excitation through its geometric interface provided by the open cavity.\\For further details on the experimental section please refer to the Appendix A.

An open cavity built on a hBN multilayer with an uniaxial indicatrix allows direct visualisation of OSHE in the system. The four-lobe structure of the S3 Stokes vector in k-space confirms a large contribution of Dresselhaus interaction at a higher momentum space.  However, in the lower momentum k-space, Rashba coupling is prominent. This prompted us to explore a unique theoretical framework to understand this phenomenon. We assume that the leaky Fabry–Perot cavity formed by the partially reflective top surface of the hBN slab and the back mirror operates well within the conventional weak coupling (Purcell) regime, satisfying $\sqrt N\,g \;\ll\; \kappa \; $  where $\sqrt N\,g$ is the collective vacuum Rabi coupling, g is the coupling between the bright exciton and the cavity mode and $\kappa$ is the cavity field decay rate. In this regime, the cavity accelerates the decay of the optically active bright exciton, rapidly depleting the bright-exciton population and transferring it to the long-lived spin-dark exciton (SDE) doublet $\{\Dket1,\Dket2\}$

In the case of $WS_2$, two SDEs are degenerated at the exciton energy $\omega_0$, lying approximately 20 to 50 meV below the bright exciton due to the conduction band spin-orbit coupling\cite{Molas2017, Zhang2017}. The SDE is formed from a conduction band electron in the lower spin sub-band and a valence band hole, which is a combination that is optically forbidden (parallel spins) and lies below the bright exciton, which is formed from the upper conduction sub-band electron. 
The OFF-resonant laser drives the SDE doublet through a second-order Schrieffer–Wolff (SW) process in which its in-plane spatial phase pattern — the phase difference accumulated by the laser wavefield across neighbouring tungsten sites on the $C_3$-symmetric W sublattice — selectively couples to the dark Fourier components of the three-site triangular model. The outcome of this coupling is twofold. First, it generates an effective Rashba Hamiltonian acting on the SDE, and secondly, it prepares the SDE doublet in a specific coherent superposition, \begin{equation}
\left|\Psi_0\right\rangle
=
\frac{1}{\sqrt{2}}
\left(
\left|D_1\right\rangle
+
e^{2i\varphi_{\mathbf{k}}}
\left|D_2\right\rangle
\right)
\end{equation}
\\
This initial state is linearly polarised — equal LCP and RCP populations with a definite relative phase set by twice the azimuthal angle of the laser’s in-plane wavevector $\varphi _{kl}$. The polarisation of the laser itself (linear, circular, or unpolarised) has no effect on this phase; only the direction $\varphi _{kl}$  of the laser tilt matters.\\
Through a SW reduction an effective Rashba Hamiltonian is obtained that acts on the SDE doublet $\{\Dket1,\Dket2\}$ at each emission wavevector $k_{EM}$:
\begin{equation}
H_{\mathrm{SOC}}=\vtri\,(k_x\tx+k_y\ty), \qquad \vtri=\frac{\Omega^2 a}{2\sqrt 3\,\Da}.
\label{eq:rashba}
\end{equation}

where $\tau_x$  and $\tau_y$ are Pauli matrices on the SDE doublet and the Rashba coupling constant is $v_\Delta$ and $\Omega$ is the laser Rabi frequency, a is  0.32 nm the nearest-neighbour W–W distance, and $\Delta_a$ is the laser-exciton detuning parameter. In principle, the coupling constant is tunable, but as we see below, the Rashba SOC is obscured by the Dresselhaus SOC of the hBN layer, making this dependence difficult to isolate. \\
The Hamiltonian in Eq.(2) splits the SDE doublet into two branches with energies
\begin{equation}
  \Epm(\kpar) = \omo + \alpha\kpar^2 \pm \vt\kpar, \qquad \kpar = \sqrt{k_x^2+k_y^2},
  \label{eq:dispersion}
\end{equation}

where $\alpha=\hbar^2/2m_X$ is the exciton kinetic mass coefficient and $\kpar = \sqrt{k_x^2+k_y^2},$ This leads to the two distinguishable parabolas for the dispersion curves for linear polarisations (TE and TM) as experimentally observed for $k_x=0$.

As the SDE at wavevector $\mathbf{k}_{\mathrm{EM}}$ decays radiatively at rate $\gamma$, the emitted photons LCP and RCP amplitudes in the frequency domain are given by a two-branch Wigner--Weisskopf calculation.

Defining the complex propagators
\begin{equation}
L_{\pm} = \gamma + i\left(\pm v_{\Delta} k_{\parallel} - \delta\right),
\end{equation}
where
\begin{equation}
\delta \equiv \omega - \omega_0
\end{equation}
for probing frequency/energy $\omega$, and the branch overlap amplitudes
\begin{equation}
c_{\pm} = \frac{1 \pm e^{-i\theta}}{2},
\end{equation}
with
\begin{equation}
\theta = \arg(k_x+i k_y)-2\varphi_{kL},
\end{equation}
we obtain the emitted field amplitudes as
\begin{equation}
A_{\mathrm{LCP}}
=
\frac{1}{\sqrt{2}}
\left(
\frac{c_{+}}{L_{+}}
+
\frac{c_{-}}{L_{-}}
\right),
\qquad
A_{\mathrm{RCP}}
=
\frac{k_x+i k_y}{k_{\parallel}}
\frac{1}{\sqrt{2}}
\left(
\frac{c_{+}}{L_{+}}
-
\frac{c_{-}}{L_{-}}
\right).
\end{equation}

The factor
\begin{equation}
\frac{k_x+i k_y}{k_{\parallel}}
=
e^{i\varphi_{\mathbf{k}}}
\end{equation}
is the photon helicity vector. As the emission direction rotates once around the origin in $k$-space, the relative phase between $A_{\mathrm{RCP}}$ and $A_{\mathrm{LCP}}$ also rotates once, encoding the winding number $w=1$ of the Rashba texture.

The three Rashba Stokes parameters before the hBN slab are then,
\begin{equation}
S_3
=
\frac{
|A_{\mathrm{LCP}}|^2-|A_{\mathrm{RCP}}|^2
}{
|A_{\mathrm{LCP}}|^2+|A_{\mathrm{RCP}}|^2
},
\end{equation}

\begin{equation}
S_1
=
\frac{
2\operatorname{Re}
\left[
A_{\mathrm{LCP}}A_{\mathrm{RCP}}^{*}
\right]
}{
|A_{\mathrm{LCP}}|^2+|A_{\mathrm{RCP}}|^2
},
\end{equation}

\begin{equation}
S_2
=
-\frac{
2\operatorname{Im}
\left[
A_{\mathrm{LCP}}A_{\mathrm{RCP}}^{*}
\right]
}{
|A_{\mathrm{LCP}}|^2+|A_{\mathrm{RCP}}|^2
}.
\end{equation}

In the absence of any birefringent optics downstream, $S_3$ would be the detected Stokes parameter. It has a pure Rashba ($w=1$) angular texture with a nodal line at angle $\varphi_k$ that rotates at twice the rate of the laser azimuth.

The emitted photon must propagate through the birefringent hBN slab (thickness $L$, ordinary index $n_o$, extraordinary index $n_e$, with $n_o>n_e$ for hBN). Because the slab's optic axis is perpendicular to the layer, photons propagating at angle $\theta$ to the normal accumulate different phases for TE and TM polarisations. In the LCP/RCP basis, this birefringence can be modelled as the Jones matrix

\begin{equation}
M_o =
\begin{pmatrix}
\cos(\Delta/2) &
i\sin(\Delta/2)e^{-2i\phi_b}
\\[4pt]
i\sin(\Delta/2)e^{2i\phi_b} &
\cos(\Delta/2)
\end{pmatrix},
\end{equation}

where, $\phi_b = \arg(k_x+i k_y)$
is the emission azimuth and the total retardance is
$\Delta = \Delta_0 + \beta k_{\parallel}^{2},$ Also, $\Delta_0 =
({2\pi L}/\lambda)(n_o-n_e)$
and $\beta =
\ [1/({2k_0})](1/n_o-1/n_e).$

The term $\Delta_0$ is the zero-order ($k$-independent) retardance set by the slab thickness and emission wavelength. The term $\beta k_{\parallel}^{2}$ carries the angular (Dresselhaus) structure through the factor $e^{\pm 2i\phi_b}$ which winds twice per revolution in $k$-space ($w=2$). Applying the Jones matrix to the Rashba amplitudes, we obtain the $S_3$ parameter of the emission exiting the hBN slab as,

\begin{equation}
S_3'(\omega,k_x,k_y)
=
\cos\Delta \cdot S_3
+
\sin\Delta
\left[
S_1\sin(2\phi_b)
-
S_2\cos(2\phi_b)
\right].
\end{equation}

This equation has a transparent geometric meaning: the hBN slab rotates the Rashba Stokes vector by angle $\Delta$ in the plane spanned by $S_3$ (circular polarisation) and the Dresselhaus projection
\begin{equation}
S_1\sin(2\phi_b)-S_2\cos(2\phi_b)
\end{equation}
($w=2$, linear-polarisation). The $\cos\Delta$ term preserves the Rashba ($w=1$) two-lobe structure while the $\sin\Delta$ term adds the Dresselhaus ($w=2$) four-lobe structure. 
\\
Several mechanisms other than the laser-induced SDE doublet Rashba can in principle generate a w=1, linearly-dispersing spin-orbit field in the $WS_2$/hBN system. However, all of them can be ruled out under the present experimental conditions. The long-range Coulomb exchange interaction between the electron and hole (excitonic Rashba) in the bright exciton can generate an effective \textit{k}-linear spin-orbit field acting on the exciton pseudospin with winding number w=1\cite{Maialle1993,Glazov2014,Yu2014}. However, this contribution is valley-antisymmetric and the S3 signals produced by the K and K’ valleys cancel under unpolarised excitation. Gmitra and Fabian  showed that proximity to transition-metal dichalcogenides induces interface Rashba SOC in adjacent 2D materials\cite{Gmitra2015}. Such an interface-induced Rashba effect acting on the $WS_2$ exciton is again valley-antisymmetric. In the strong coupling regime, the combination of the photonic TE-TM splitting of the cavity mode and the excitonic longitudinal-transverse exchange splitting produces an effective Rashba-Dresselhaus SOC acting on the polariton pseudospin\cite{Spencer2021,LiangJ2024, Kavokin2004}.  The laser-induced SDE doublet Rashba mechanism proposed here survives because it is driven by the spatial phase pattern of the laser wavefield across the W sublattice rather than by valley polarisation. 

\begin{figure}[h]
    \centering
    \includegraphics[
        width=0.95\linewidth
    ]{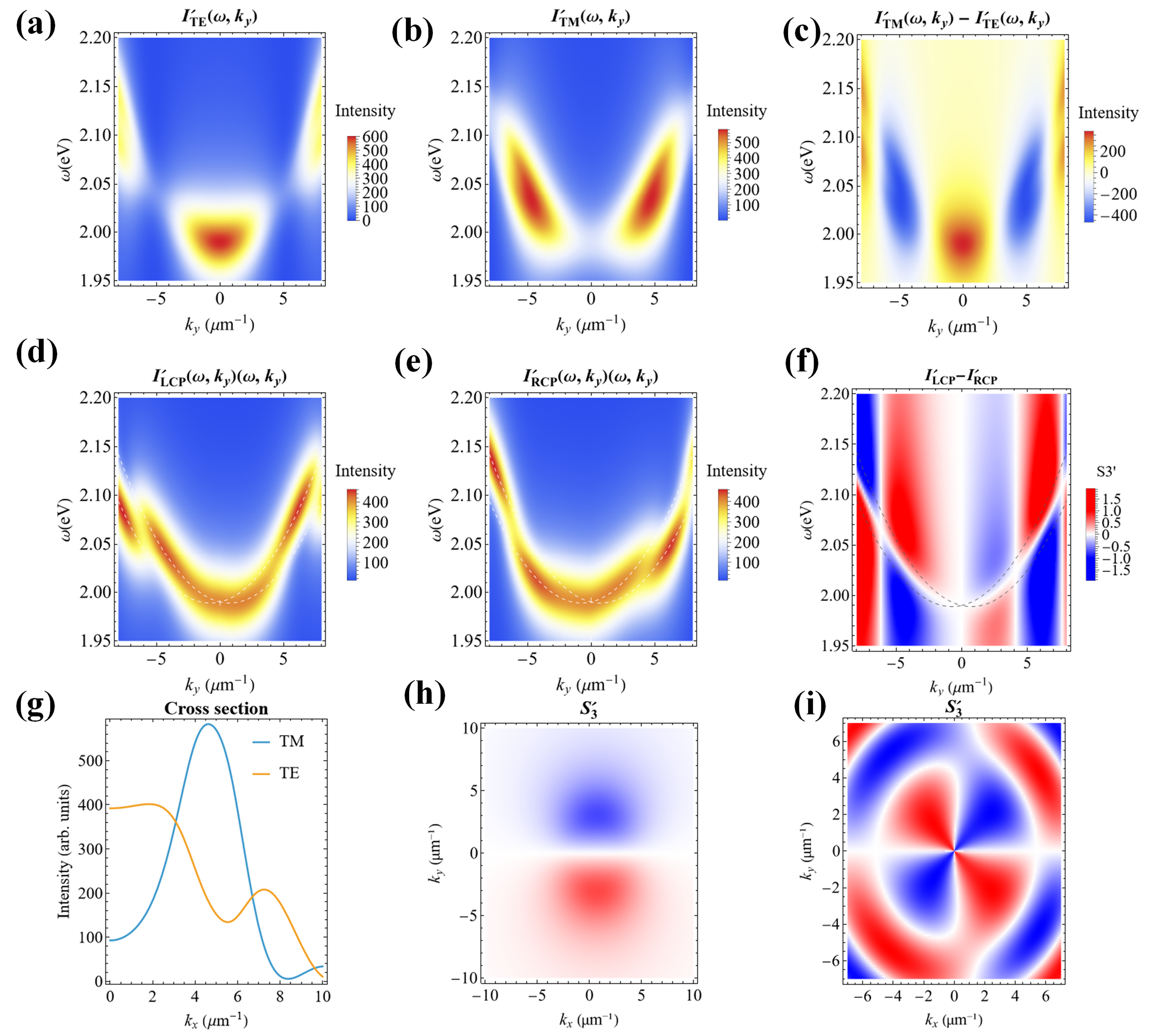}

    \caption{
        Numerical computation of the dispersions and $S_3'$: The first row (a, b, c) shows the TE and TM dispersion at $k_x$ = 0 and the difference between them (compare with Figs. 2 and 4). The second row (d, e, f) shows the LCP and RCP dispersion and their difference for $k_x$ = 0 (compare with Fig 3). In the third row (g, h, i), the first panel is the cross-section for TE, TM dispersions ($\omega=2.0$ eV ). The second and third panels show $S_3'$ vector as a function of $k_x$ and $k_y$ with and without the hBN slab, respectively ($\omega=2.0$ eV ). 
    }

\end{figure}

The TE, TM, LCP, and RCP dispersion curves, as well as the angle-resolved two-dimensional map of $S_3'(\omega,k_x,k_y)$ as a function of $k_x$ and $k_y$, were computed numerically based on the theory presented above. For these calculations, we used the following dispersion relation in Eq.~(3):
\begin{equation}
E_{\pm}(k_{\parallel})
=
\omega_0
+
\alpha k_{\parallel}^{2}
\pm
\hbar\nu_{\Delta} k_{\parallel},
\end{equation}
with $\omega_0 = 1.99~\mathrm{eV}$, $\alpha = 2\times10^{5}~\mathrm{eV\,\text{ \AA}^2}$, and $\hbar\nu_{\Delta} = 30~\mathrm{eV\,\text{ \AA}}$. The laser azimuthal angle was set to $\phi_{kL}=0$. The radiative linewidth, consistent with the experimental observations, was taken as $\hbar\gamma = 0.035~\mathrm{eV}$, which is close to other independently reported values in the literature\cite{Cadiz2017,Ajayi2017}.

The birefringence of hBN was taken from reported values, with $\Delta n = 0.399$, and the effective optical path length through the hBN slab was fitted to obtain agreement with the experimental $S_3'$ vector. Note that the Dresselhaus contribution is dominant, whereas the Rashba contribution is comparatively weak, making it difficult to directly verify the dependence of the Rashba term on parameters such as the laser intensity and detuning. Nevertheless, the Rashba splitting is essential for generating a nonzero $S_3$ component, which constitutes a significant contribution to the $S_3'$ signal observed after propagation through the hBN slab.

The numerical results are shown in Fig.5. We find excellent agreement between the numerical calculations and the experimental observations, particularly for the two-dimensional map of $S_3'$ and the measured dispersion curves.
\\Additional information and detailed theoretical calculation can be found at Appendix B.

\section*{CONCLUSION:}
In conclusion, we have demonstrated OSHE in open optical cavities driven by dark strong coupling. The OSHE effect in this system has been induced by the combination of: (1) an effective magnetic field generated by Dresselhaus, and (2) the Rashba spin-orbit interactions. The emergence of the Dresselhaus coupling has been governed by the out-of-plane anisotropy of the bare hBN. OSHE is observed in both reflection and emission measurements, as confirmed by the shift of the parabolic dispersion band under the selected spin polarisation. On top of that, in the $k_x$ versus $k_y$ plot, a four-lobe structure has emerged clearly, further supporting our claim. A detailed theoretical calculation based on the SDE allows the system to undergo symmetry breaking, thereby inducing Rashba coupling. Large birefringence in the hBN slab amplifies the SOC. Apart from these, we have shown the generation of linear and circularly polarised emission from an unpolarized excitation source in open cavities. This simple platform may have direct application in polarisation-sensitive photonic devices. 
\section*{MATERIALS AND METHODS:}
\subsection*{Sample Fabrication:}
The heterostructure sample consists of an Ag mirror at the bottom, followed by hBN and $WS_2$ monolayers placed directly on top of the mirror. The Ag mirror is created by sputtering 120 nm of silver onto a 1mm-thick BK7 glass substrate. Ultrasonication is done to optically clean the substrate using distilled water and ethanol. The thin-film deposition process is performed using a BT300 plasma sputterer from HHV. The hBN and $WS_2$ crystals were purchased from HQ Graphene Inc. The layers are manually exfoliated in a PDMS substrate using the scotch tape method.  The thickness of the hBN was then determined from the optical interference data in the reflection measurement of bare hBN. The $WS_2$ monolayer can be easily visualised by its characteristic emission at 614 nm under 532 nm diode-laser excitation. To fabricate the heterostructure, hBN and $WS_2$ were transferred onto the Ag substrate one at a time using a custom-built transfer stage. First, hBN is transferred onto the Ag substrate, followed by the $WS_2$ monolayer. For the reflection studies, the bare hBN has been directly transferred onto the glass substrate.
\subsection*{Polarisation-resolved reflection and k-space emission mapping}: The setup includes a Nikon Ti2-Eclipse inverted microscope with halogen lamp illumination, an HR300 spectrometer from Princeton Instruments, a SuperK laser from NKT Lasers, and other optical components from Thorlabs. A 100X objective of 0.8 NA has been used in all experiments. The fabricated sample was placed on the microscope's sample stage, and all measurements were performed in reflection mode. We used a 4f-relay optical setup to propagate light from the microscope to the spectrometer. For Fourier-space imaging, we used a convex lens as the back focal plane lens immediately before the spectrometer slit. To resolve the components of linear polarisation, we have used a linear polariser in the analyser path, whereas for circular polarisation, we have used a quarter-wave plate and a linear polariser.

\newpage

\section*{AUTHOR INFORMATION}
\subsection*{Corresponding Authors }
Prof Anil Shaji - School of Physics, Indian Institute of Science Education and Research (IISER) Thiruvananthapuram, Vithura, Kerala, India – 695551; Email:shaji@iisertvm.ac.in\newline
Dr Jino George – Indian Institute of Science Education and Research (IISER), Mohali, Punjab 140306, India; Email: jgeorge@iisermohali.ac.in; orcid.org/0000-0002-3558-6553
\subsection*{Authors}
Kunal Banerjee – Indian Institute of Science Education and Research (IISER), Mohali, Punjab 140306, India.\newline
Dr Ben Johns – Indian Institute of Science Education and Research (IISER), Mohali, Punjab 140306, India. \newline
(Current Address: Ultrafast Nanoscience Unit, Department of Physics, Umeå University, Linnaeus väg 24, 90187 Umeå, Sweden; Umeå Centre for Microbial Research, Umeå, 901 87, Sweden.); orcid.org/0000-0002-6752-5931.\newline
Nitin Yadav – Indian Institute of Science Education and Research (IISER), Mohali, Punjab 140306, India; orcid.org/0009-0003-6190-5604.

\section*{ACKNOWLEDGMENT}
J.G. thank the ANRF-core research grant (CRG/2023/001122) and IISER Mohali for the institute start-up grant. AS acknowledges the support of the National Quantum Mission, Government of India, through a grant from the Foundation for Quantum Computing Innovation. K. B. and N. Y. thank IISER Mohali for the research fellowship. 
\newpage

\section*{ Appendix A : Experimental Procedure and Results}
\subsection*{Experimental Set up}

The experimental setup includes an excitation laser source, a detection spectrometer, and other optical components. An optical microscope is used to probe the small monolayer region with a 100x magnification objective of numerical aperture of 0.8. 

\begin{figure}[H]
    \centering
   \includegraphics[width=0.9\textwidth]{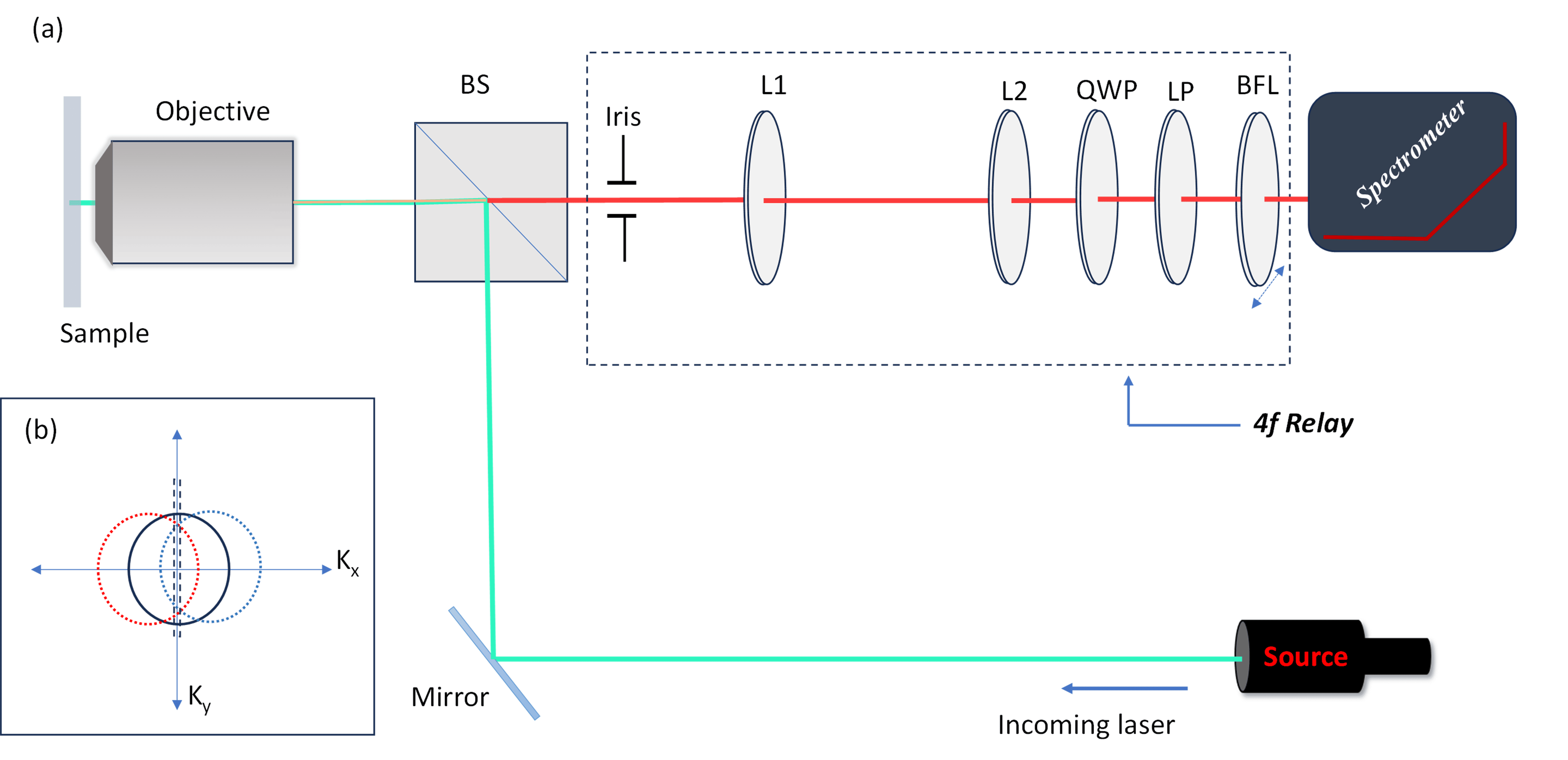}
    \caption*{
        \textbf{Figure A1:} (a) Schematic of the polarisation-resolved 3D tomographic Fourier plane imaging set-up. (b) By translating the Back Focal Lens (BFL) in the detection path, we can scan the various $k_{x}$ values other than $k_{x}$ = 0
    }
    
\end{figure}

For emission studies, we have used a SuperK laser from NKT Photonics. Its wavelength can be varied from 500 to 800 nm. A 550 nm long-pass filter is placed in the emission path to prevent the laser from reaching the spectrometer. To transport the beam from the microscope to the spectrometer, we have used a 4f relay lens setup. Just before the spectrometer slit, a BFL is placed to convert the real-space spectra to Fourier space for angle-dependent studies. For the measurement in the linear polarisation basis, only a linear polariser is used in the detection path, while for the circular basis, a combination of a linear polariser and a quarter-wave plate is used. Data has been recorded for multiple angles of the Quarter Wave Plate and Linear Polariser at 10-degree intervals. K-space tomography has also been studied using lateral translation of the Back Focal Lens. In Figure A1(b), we show that multiple $k_{x}$ spots can be measured using this technique.


\subsection*{TM-TE splitting in bare h-BN}

\begin{figure}[H]
    \centering
    \includegraphics[width=0.9\textwidth]{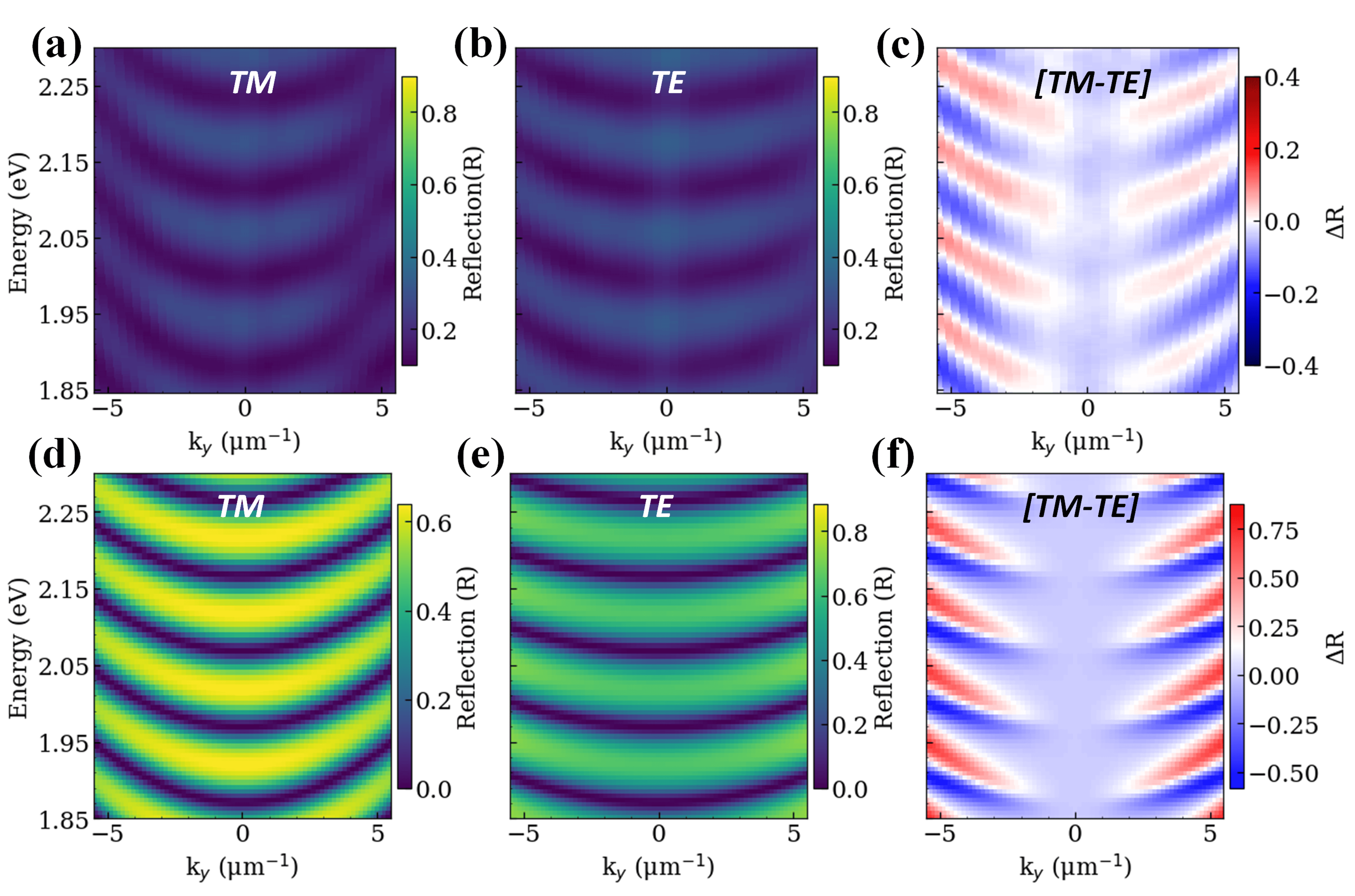}
    \caption*{
        \textbf{Figure A2}: Angle-resolved TM-TE reflection spectra for Bare h-BN of thickness 2290 nm. (a, b, c) Experimental data for the TM, TE component and TM-TE difference spectra. (d, e, f) TMM simulated data for the TM, TE component and TM-TE difference spectra.
    }
    
\end{figure}
The optically confined modes of h-BN are studied using reflection micro-spectroscopy. Mainly, the Transverse Magnetic and the Transverse Electric modes are observed carefully to understand their role in our experiment. It is found that the TM TE modes significantly differs from each other in the higher momentum values. This imbalance between the TM-TE modes is one of the major contributing factors to generate spin orbit interaction in open cavities.\newline
In the experiment, a back-focal setup is used to capture the angle-resolved evolution of TM-TE modes. TM modes are found to be growing more rapidly in energy with increasing angle than the TE modes, which is consistent through out different experiments in open cavity.\newline
The simulated results are also found to show the similar trend like the experimental values. Simulations are done using Transfer Matrix Method(TMM) calculations considering an out-of-plane anisotropy in the h-BN crystal.

\newpage


\subsection*{Demonstration of spin orbit interaction effects in open cavity(h-BN thickness 1355nm)}

\begin{figure}[H]
    \centering
    \includegraphics[width=0.9\textwidth]{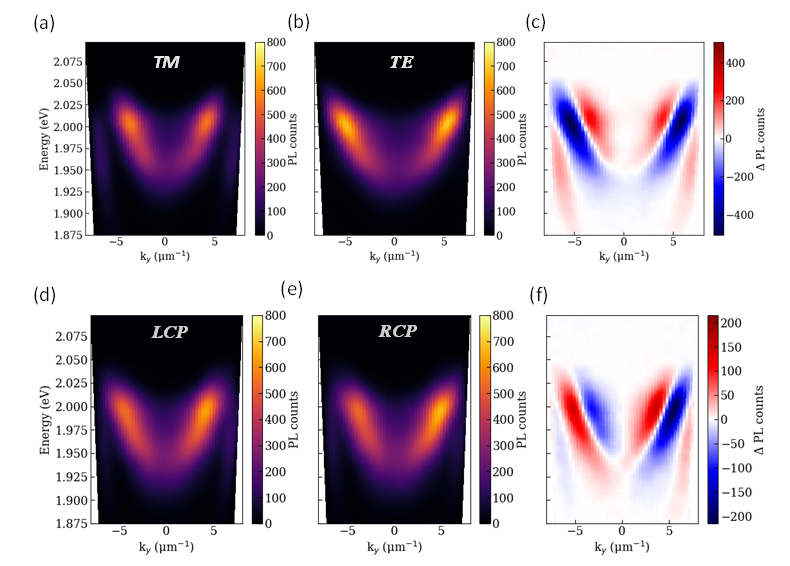}
    \caption*{
       \textbf{Figure A3}: Angle-resolved emission data for open cavity (a, b, c) TM, TE component and the TM-TE difference spectra. (d, e, f) LCP, RCP component and LCP-RCP difference plot.
    }
  
\end{figure}
We have conducted similar linear polarisation and circular polarisation studies on the Ag-hBN-$WS_2$ heterostructure with an hBN thickness of 1355 nm to confirm the generality of the experiment across various hBN thicknesses. In linear polarisation, we found that the TE mode spread wider in the momentum space than the TM mode. TM mode reaches similar energies to TE at much lower momentum values. In the circular polarization, we can see a clear separation of spin-split LCP and RCP bands in the momentum space. This is visualised better through the difference plot.

\newpage

\subsection*{Periodic changes in the dispersion width}

\begin{figure}[H]
    \centering
    \includegraphics[width=0.9\textwidth]{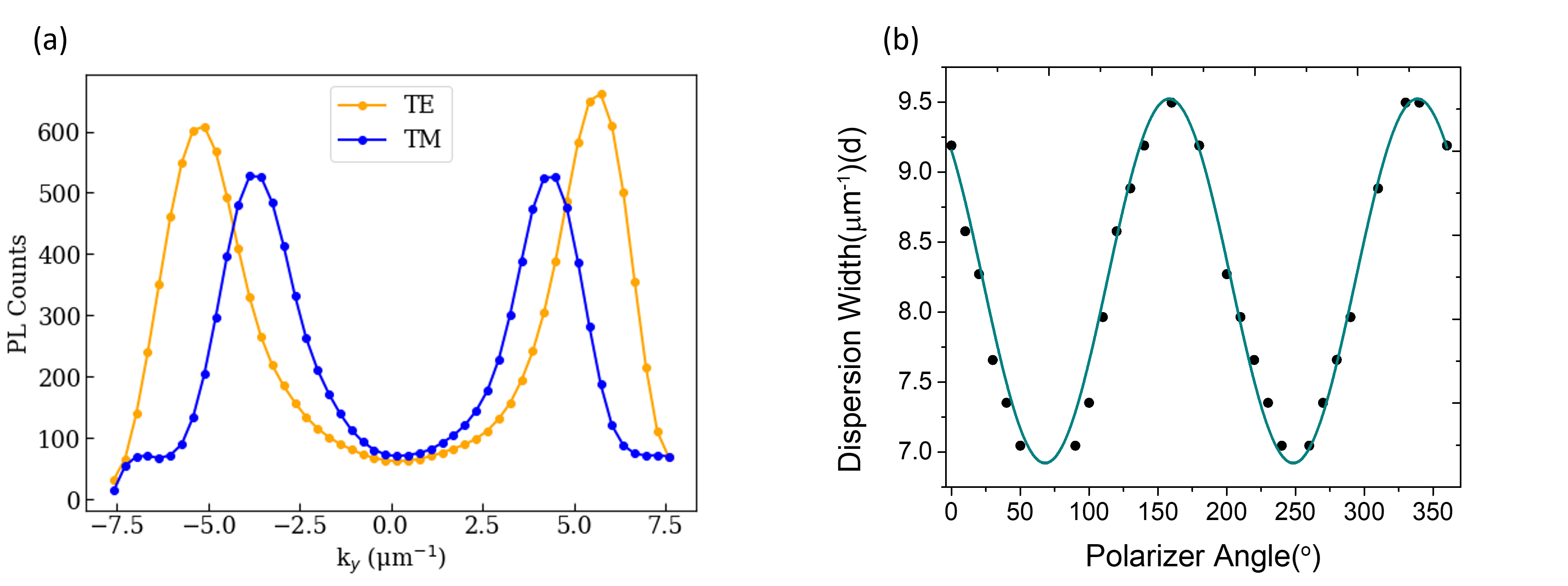}
    \caption*{
        \textbf{Figure A4}: (a) Momentum cross-section of TE and TM dispersion spectra along a fixed energy of 1.99 eV. (b) The dispersion width modulation with other variable linear polarisation angles.
    }
    
\end{figure}

When we take a cross-section of the energy-momentum dispersion of polariton emission at a fixed energy, we obtain a double-peak profile centred at zero momentum. We calculated the difference between the two peaks' momentum values and defined it as the dispersion width, d. Figure A4(a) shows the profiles of the TE and TM components of the dispersion shown in Figure A3(a, b). We also measured the variation in dispersion with polariser rotation at 10-degree intervals. We found that the dispersion width repeatedly changed as we rotated the polariser. So, we fit it with a sine curve and eventually found the periodicity to be 180 degrees. 

\newpage


\subsection*{Generation of linearly polarized modes at higher momenta}

\begin{figure}[H]
    \centering
    \includegraphics[width=0.9\textwidth]{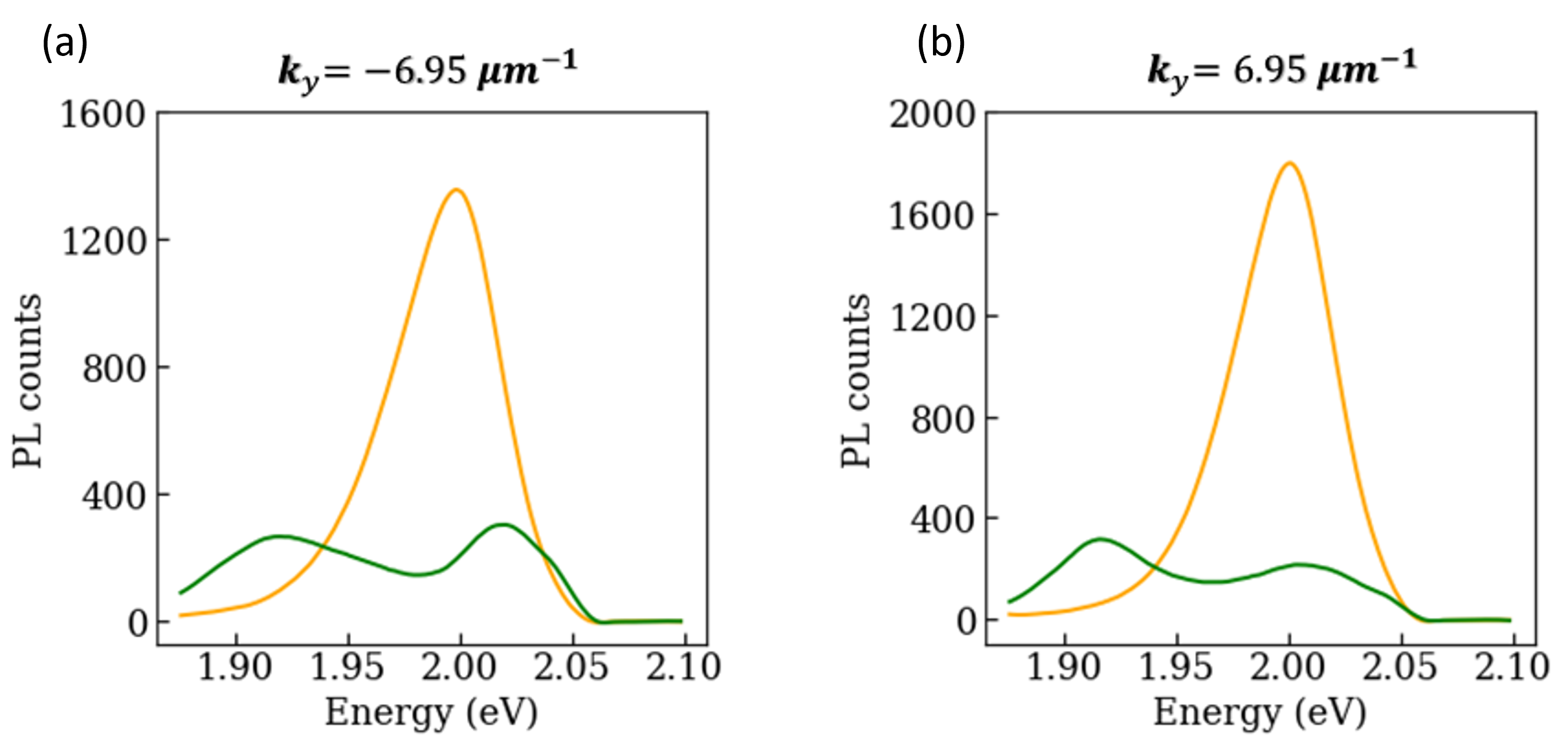}
    \caption*{
       \textbf{Figure A5}: Emission spectra of TE(Green)and TM(Orange) modes of Ag-hBN-$WS_2$ heterostructure for hBN thickness 1460 nm at specific momentum values, (a) -6.95 $\mu\mathrm{m}^{-1}$ and (b) $6.95 \mu\mathrm{m}^{-1}$  
    }
    
\end{figure}

We have extracted the dispersion emission spectra by taking cross-sections at momentum values of -6.95 $\mu m^{-1}$ and 6.95 $\mu m^{-1}$ for both TE and TM modes, as shown in figures 5(a) and 5(b). It is found that one mode emits prominently, while the other is suppressed. As a result, we obtain linearly polarised light with a high extinction coefficient. This effect can be attributed to the higher curvature of the TM mode parabolic dispersion. Due to this, it touches the exciton energy value quickly and opens up the cavity channel for emission. While TE mode is wider and spread in higher k-space, that’s why it shows a very tiny satellite emission at lower energies in this case. The degree of linear polarisation is calculated from this, which turns out to be 74 \%.

\newpage


\subsection*{Generation of circularly polarized modes at higher momenta}

\begin{figure}[H]
    \centering
    \includegraphics[width=0.9\textwidth]{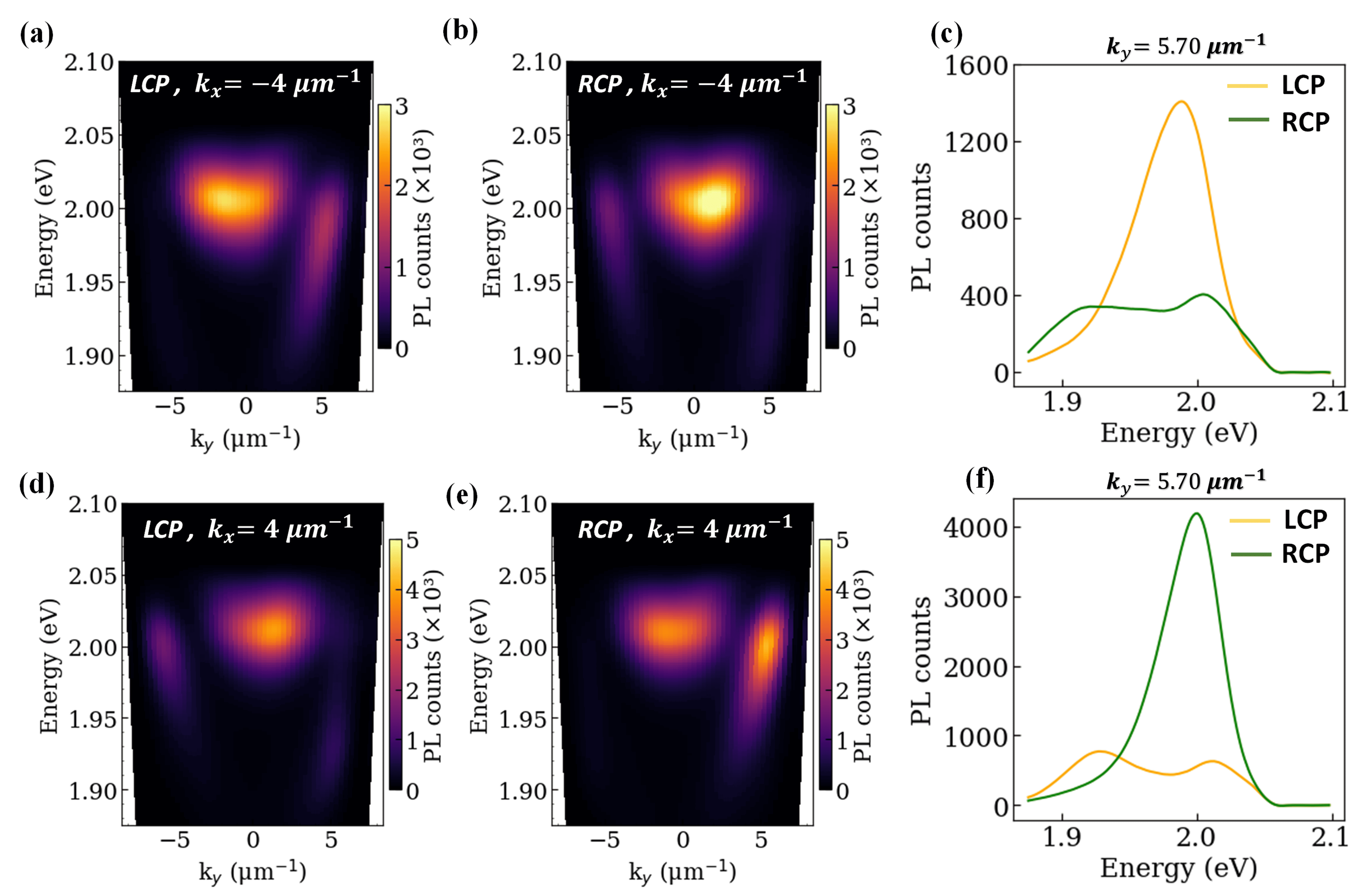}
    \caption*{
        \textbf{Figure A6}:  Dispersive emission from Ag-hBN-$WS_2$ open cavity for hBN thickness 1460 nm. (a) LCP component at $k_{x}$ = -4 $\mu\mathrm{m}^{-1}$ , (b) RCP component at $k_{x}$ = -4 $\mu\mathrm{m}^{-1}$ , (d) LCP component at $k_{x}$  = 4 $\mu\mathrm{m}^{-1}$ , (e) RCP component at $k_{x}$ = 4 $\mu\mathrm{m}^{-1}$ , (c) PL spectra at $k_{y}$ = 5.70 $\mu\mathrm{m}^{-1}$  of LCP and RCP components at $k_{x}$= -4 $\mu\mathrm{m}^{-1}$, (f) PL spectra at $k_{y}$ = 5.70 $\mu\mathrm{m}^{-1}$  of LCP and RCP components at $k_{x}$ = 4 $\mu\mathrm{m}^{-1}$. 
    }
    
\end{figure}

The k-space tomographic measurement is performed by varying the position of the BFL. The BFL is adjusted to cover the higher k-space by probing larger $k_x$ values. At these $k_x$ values, the components of dispersive emission are resolved into LCP and RCP counterparts. It has been found that at higher $k_y$, circularly polarised luminescence dominates. The polarisation properties completely reversed when the measurement was done at positive $k_x$ values compared to negative $k_x$ values. In Figure A6, it can be clearly seen that LCP emission is dominated at a positive $k_y$ when we measure at a positive $k_x$ and \textit{vice versa}. A completely opposite trend is observed when measured at a negative $k_x$ value. So, this experiment confirms that our system is capable of producing high circularly polarized emission at higher momentum.

\newpage

\subsection*{Parabolic fitting of TM-TE splitting in bare h-BN}

\begin{figure}[H]
    \centering
    \includegraphics[width=0.85\textwidth]{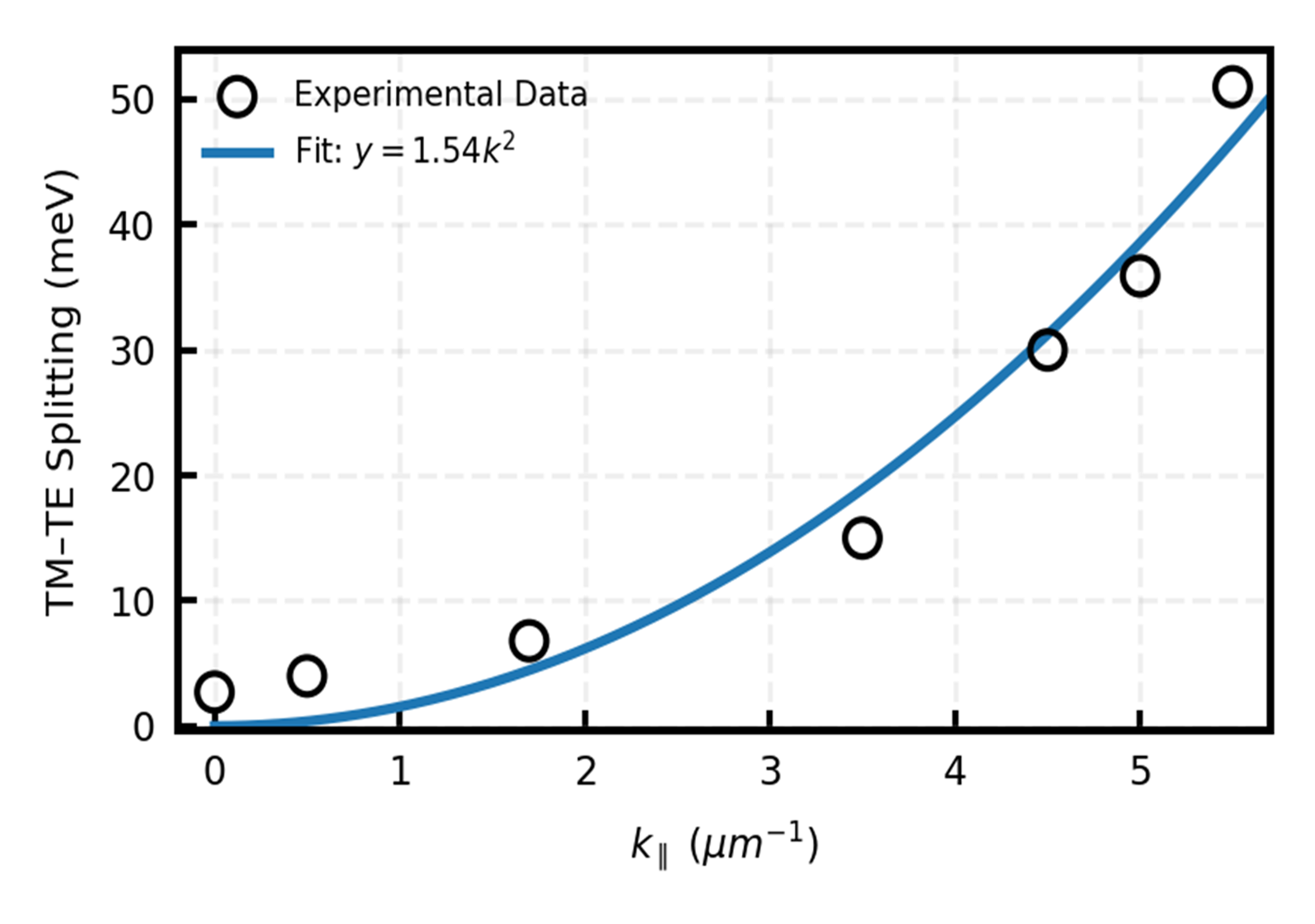}
    \caption*{
        \textbf{Figure A7}: Experimental TM-TE splitting variation with in plane momentum for h-BN thickness of 2290 nm and its quadratic fit.
    }
    \label{fig:S1}
\end{figure}

The experimental values of TM-TE splitting is fitted with a parabolic curve of $y=\beta x^2$. The strength of TM-TE splitting, is found to be $\beta$ = $1.54 \times 10^5$ eV.$\text{ \AA}^{2}$ which is considerable to generate effective spin orbit interaction without any complex heterostructure.

\newpage

\subsection*{Calculation of Rashba coupling strength}

\begin{figure}[H]
    \centering
    \includegraphics[width=0.7\textwidth]{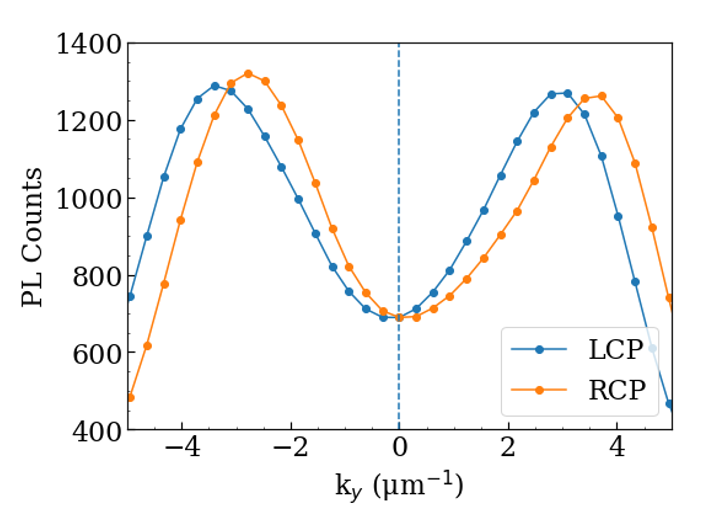}
    \caption*{
       \textbf{Figure A8}: The cross-section of LCP and RCP dispersion curves of emission for open cavity (hBN thickness 1460 nm) at 2.01 eV. 
    }
   
\end{figure}

The figure clearly illustrates a relative shift in dispersion between LCP and RCP. The high-intensity peaks correspond to two of the primary lobes of the dispersion. The peak difference between this dispersion is 0.619 $\mu m^{-1}$. \\
The formula for the dispersion shift between two parabolas is,

\begin{equation}
\Delta k_y = \frac{4m\alpha}{\hbar^2}
\end{equation}

Therefore, the Rashba coupling strength $\alpha$ is given by

\begin{equation}
\alpha = \frac{\hbar^2 \Delta k_y}{4m}.
\end{equation}

For the measured separation between the LCP and RCP dispersion,

\begin{equation}
\Delta k_y = 0.619~\mu\mathrm{m}^{-1},
\end{equation}

and the effective polariton mass for the calculation is taken as,

\begin{equation}
m_p = 5\times10^{-3}m_e.
\end{equation}

Using

\begin{equation}
\frac{\hbar^2}{m_e}
= 0.07620~\mathrm{meV\,\mu m^2},
\end{equation}

we obtain

\begin{align}
\alpha
&= \frac{\hbar^2\Delta k_y}{4m_p} \\
&= \frac{
(0.07620~\mathrm{meV\,\mu m^2})
(0.619~\mu\mathrm{m}^{-1})
}{
4(5\times10^{-3})
} \\
&= 2.358~\mathrm{meV\,\mu m}.
\end{align}

Thus, Rashba coupling strength is

\begin{equation}
\boxed{
\alpha \simeq 2.36~\mathrm{meV\,\mu m}
}
\end{equation}

or equivalently,

\begin{equation}
\boxed{
\alpha \simeq 23.6~\mathrm{eV\,\text{ \AA}}
}.
\end{equation}

\newpage


\section*{Appendix B : Theoretical model of the system}

The system of interest is a vertical stack that consists of a perfectly reflecting mirror, a hBN slab of various thickness above 1100 nm, and a monolayer of $\mathrm{WS_2}$ exposed at the top surface. The hBN slab together with the back mirror forms a half-open optical cavity for which the bottom boundary is a high-reflectivity mirror, while the top boundary (the hBN/air or hBN/$\mathrm{ WS_2 }$/air interface) provides only weak index-contrast reflectivity, so the cavity is strongly \emph{leaky}. The structure is excited by an unpolarised continuous-wave laser at $\lambda_L=532\,$nm.

For any thickness of the hBN cavity the field maxima or antinode always found to be at the top of the h-BN that coincides with the $\mathrm{WS_2}$ layer at the top. This maximises the local field amplitude seen by the excitons in the WS$_2$ layer and the excitons can be coupled to the cavity.  The full physical picture we develop has four components/assumptions which are listed below.
\begin{enumerate}
\item \textbf{Strong coupling at the antinode:} Because the $\mathrm{WS_2}$ A-excitons sit at a field antinode, they couple strongly to the cavity mode. The relevant collective object is the totally symmetric (``bright'') exciton mode, which carries the entire collective dipole and couples to light with a superradiantly enhanced strength~\cite{Dicke1954,Gross1982}.

\item \textbf{Leaky-cavity depletion of the bright mode:} A very leaky cavity provides a fast, Purcell-enhanced~\cite{Purcell1946} loss channel for the bright mode. The bright state population leaks away on a timescale much shorter than the intrinsic dark-state lifetime, leaving the system in the subradiant \emph{dark} manifold. This justifies taking a dark state as the initial condition for the subsequent slow dynamics.

\item \textbf{Dark-state texture and SOC:} The dark states are not featureless and instead they carry a definite internal phase winding (a $C_3$ ``angular momentum'' $\ell=\pm 1$). When an obliquely incident laser imprints an in-plane momentum $\bm{k}_\parallel$, this phase texture couples to $\bm{k}_\parallel$ and produces an effective Rashba-type spin-orbit coupling of the two-dimensional dark-exciton pseudo-spin~\cite{Bychkov1984}. The resulting momentum-dependent splitting and pseudo-spin winding contributed to the optical spin-Hall effect~\cite{ Kavokin2005, Leyder2007} in the emitted light.

\item \textbf{Angle resolved photoluminiscence signatures.} The intrinsic Desselhaus SOC of the hBN layer adds to the Rashba SOC to generate the observed dispersion curves and polarisation maps. 
\end{enumerate}

An important consideration is the precise role of the cavity. While the cavity is essential for \emph{preparing} the dark state (step ii) and for \emph{amplifying} the drive (step i), the effective Rashba coupling of the dark pseudo-spin turns out to be \emph{independent of the cavity coupling $g$} at leading order.

\subsection*{Tavis-Cummings Model for Three Excitons}
\label{sec:TC}

We model the excitons coupled to the leaky cavity using the Tavis-Cummins Hamiltonian for $N$ identical two-level excitons (transition frequency $\omega_a$) coupled to a single cavity mode (frequency $\omega_c$). In the rotating-wave approximation the Tavis-Cummings (TC) Hamiltonian is~\cite{TavisCummings1968,TavisCummings1969}
\begin{equation}
H = \omega_c\,\adag\ah + \frac{\omega_a}{2}\sum_{p=0}^{N-1}\siz{p} + g\sum_{p=0}^{N-1}\left(\adag\,\sm{p} + \ah\,\sip{p}\right),
\label{eq:TC-lab}
\end{equation}
with $\hbar=1$. Here $\adag,\ah$ are the cavity creation/annihilation operators, $\sip{p},\sm{p},\siz{p}$ are the Pauli raising/lowering and inversion operators of exciton $p$, and $g$ is the single-exciton vacuum Rabi coupling (assumed equal for all excitons because they sit at a common antinode). The three terms are the cavity energy, the exciton energy, and the dipole (Jaynes-Cummings type) exciton-photon coupling~\cite{JaynesCummings1963}.

Photon loss through the leaky top mirror and non-radiative/free-space exciton decay are described by a Lindblad master equation,
\begin{equation}
\dot{\rho} = -i[H,\rho]
 + \kappa\,\mathcal{D}[\ah]\rho  + \gamma\sum_{p}\mathcal{D}[\sm{p}]\rho, \qquad \mathcal{D}[L]\rho = L\rho L^\dagger - \tfrac12\{L^\dagger L,\rho\},
\label{eq:lindblad}
\end{equation}
where $\kappa$ is the cavity (photon) decay rate and $\gamma$ the single-exciton decay rate~\cite{Breuer2002}. The leaky (open) cavity corresponds to a large value of $\kappa$.

\subsubsection*{Collective (Fourier) modes and the bright/dark split}
\label{sec:fourier}

The coupling in Eq.~\eqref{eq:TC-lab} involves only the totally symmetric combination of excitons. We define the discrete Fourier (collective) operators
\begin{equation}
\Tp{q} = \frac{1}{\sqrt{N}}\sum_{p=0}^{N-1}\wt^{\,qp}\,\sip{p}, \qquad
\Tm{q} = (\Tp{q})^\dagger = \frac{1}{\sqrt{N}}\sum_{p=0}^{N-1}\wt^{\,-qp}\,\sm{p},
\label{eq:fourier-def}
\end{equation}
with the cube root of unity $\wt = e^{-2\pi i/3}$. The $q=0$ mode $\Tp{0}=\frac{1}{\sqrt N}\sum_p\sip{p}$ is the \emph{bright} (superradiant) mode and we can rewrite the Travis-Cummings coupling term as
\begin{equation}
g\sum_p(\adag\sm{p}+\ah\sip{p}) = \sqrt{N}\,g\,(\adag\Tm{0}+\ah\Tp{0}).
\label{eq:bright-coupling}
\end{equation}
This means that the photon couples only to $q=0$ with collective strength $\sqrt N g$. The modes $q\neq 0$ are \emph{dark} and they have no dipole matrix element with the cavity and are therefore subradiant. In other words, without any other perturbations or imperfections in the cavity, the dark states cannot couple to the electromagnetic field and decay. 

\subsubsection*{Three regimes}
\label{sec:regimes}

The single relevant collective coupling is $g_N\equiv\sqrt N g$. With cavity loss $\kappa$ and exciton loss $\gamma$, we can identify the following three standard regimes:
\begin{itemize}
\item \textbf{Strong coupling:} $g_N \gg \kappa,\gamma$. The bright exciton and the photon hybridise into two resolved polariton branches split by $2g_N$. In this regime the vacuum Rabi oscillations are visible and this feature can be spectrally resolved. Since vacuum Rabi splitting is not experimentally observed, we are not working in this regime. \item \textbf{Weak coupling / Purcell:} $g_N \ll \kappa$ (bad cavity). No Rabi
splitting; instead the bright mode acquires a Purcell-enhanced decay
\begin{equation}
\Gamma_{\mathrm{bright}} = N\gamma\left(1+C\right), \qquad C \equiv \frac{4 g_N^2}{\kappa\,\gamma}=\frac{4Ng^2}{\kappa\gamma},
\label{eq:purcell}
\end{equation}
where $C$ is the cooperativity~\cite{Purcell1946}. The factor $N$ in front is the superradiant enhancement of the bright collective dipole. This is the regime that is most relevant to the current experiment. 
\item \textbf{Dark mode:} the dark modes do not couple to $\ah$ and therefore decay only at the bare rate $\gamma$ (plus any free-space/dephasing channel), independent of $\kappa$ and $g$.
\end{itemize}

\noindent {\bf The useful window:} For rapid, cavity-enhanced cleanup of the bright mode while keeping the dynamics free of confusing polariton structure we want simultaneously
\begin{equation}
\sqrt N\,g \;\ll\; \kappa \;\ll\; \frac{4Ng^2}{\gamma} \quad \Longleftrightarrow \quad \text{(no Rabi splitting)}\ \text{and}\ (C\gg1),
\label{eq:window}
\end{equation}
which is open whenever $g\gg\gamma$ (a good emitter in a bad cavity). In this window the bright mode leaks out at $\Gamma_{\mathrm{bright}}=N\gamma(1+C)\gg N\gamma$, i.e. much faster than in free space, while the dark mode persists at $\sim\gamma$. 

\subsection*{The minimal three-site model}
\label{sec:triangle}

Monolayer $\mathrm{WS_2}$ is a direct-gap semiconductor of the $2H$ transition-metal dichalcogenide family, with the band extrema at the $K$ and $K'$ corners of the hexagonal Brillouin zone~\cite{Mak2010, Splendiani2010}. The lowest bright excitonic resonance (the A-exciton) is built from the spin-split valence band and the conduction band at $K/K'$, and is governed by the strong Coulomb interaction characteristic of two-dimensional semiconductors~\cite{Wang2018,Mak2018}. The A-exciton transition dipole lies in the monolayer plane, and the $K$/$K'$ valleys couple to left- and right-circularly polarized light, respectively (valley-selective circular dichroism)~\cite{Cao2012,Mak2012,Xu2014}. The exciton center-of-mass momentum $\bm{k}_\parallel$ is set, in an absorption process, by the in-plane wave vector of the exciting field.

The relevant electronic symmetry is set by the underlying lattice. Monolayer $\mathrm{WS_2}$ crystallizes in the 2H structure, with the W and S atoms arranged on a honeycomb lattice possessing the point group symmetry $D_{3h}$. Within this structure, the W sub-lattice alone forms a triangular lattice with three-fold ($C_3$) rotational symmetry about each W site. Because the A-exciton wavefunction is built predominantly from W $d$-orbitals, the local symmetry environment seen by an exciton is, to leading order, the $C_3$ symmetry of its three nearest W neighbors arranged at the vertices of an equilateral triangle. This three-fold symmetry — rather than any specific exciton fine-structure mechanism — is what organizes the structure of the bright/dark manifold into a totally symmetric (bright) combination and a two-dimensional dark doublet related by the $C_3$ rotation.

In our model we replace the continuum of W sites by three equivalent A-exciton emitters located at the vertices of an equilateral triangle so that we have a simple and analytically tractable model. This choice retains the $C_3$ rotational symmetry of the lattice (which controls the structure of the dark manifold) while making all collective modes and matrix elements computable in closed form. Let the three sites be labeled $p=0,1,2$, placed at
\begin{equation}
\bm{r}_p = a_0\left(\cos\theta_p,\ \sin\theta_p\right), \qquad \theta_p = \frac{2\pi p}{3}+\theta_0,
\label{eq:positions}
\end{equation}
with $a_0$ the circumradius and $\theta_0$ a fixed orientation offset. The nearest-neighbor separation (triangle side) is $a=\sqrt{3}\,a_0$. The origin of coordinates is the centroid of the triangle, so that the $C_3$ rotation $\Rop$ about the centroid permutes the sites cyclically, $p\to p+1\ (\mathrm{mod}\,3)$. Specifically if we set $\theta_0 = \pi/2$, the excitons are located at $(0,a_0)$, $(-\sqrt{3}a_0/2, -a_0/2)$ and  $(\sqrt{3}a_0/2, -a_0/2)$ (one exciton along the y axis and the other two located symmetrically on either side of the y-axis slightly below the origin).

Each site carries a two-level exciton. The bare Hamiltonian (before the laser) is the $N=3$ Tavis-Cummings Hamiltonian of Eq.~\eqref{eq:TC-lab},
\begin{equation}
H_0 = \Hcav + \Hatm + \HTC, \quad \Hcav=\omega_c \adag\ah,\quad \Hatm=\frac{\omega_a}{2}\sum_{p=0}^{2}\siz{p}, \quad \HTC=\sqrt{3} g (\adag\Tm{0}+\ah\Tp{0}),
\label{eq:H0-lab}
\end{equation}
where the last equality uses Eq.~\eqref{eq:bright-coupling} with $N=3$. The cavity couples only to the bright mode $\Tp{0}$. Each term is: the cavity photon energy ($\Hcav$); the total exciton energy, counting $+\tfrac{\omega_a}{2}$ for each excited site and $-\tfrac{\omega_a}{2}$ for each ground site ($\Hatm$); and
the collective dipole coupling of the bright exciton mode to the cavity field ($\HTC$).

\subsubsection*{Laser drive, interaction picture, and conserved quantities}
\label{sec:drive}


An external classical laser of frequency $\omega_L$ and in-plane wave vector $\bm{k}_\parallel$ drives the excitons through their dipoles. In the rotating-wave approximation the drive term is
\begin{equation}
H_{\mathrm{drive}}(t)=\frac{\Omega}{2}\sum_{p=0}^{2} \left(e^{-i\omega_L t}\,e^{\,i\bm{k}_\parallel\cdot\bm{r}_p}\,\sip{p} +\mathrm{h.c.}\right),
\label{eq:drive-lab}
\end{equation}
where $\Omega$ is the (single-site) Rabi frequency proportional to the local field amplitude at the emitter and to the transition dipole. The site-dependent phases $\phi_p\equiv\bm{k}_\parallel\cdot\bm{r}_p$ encode the obliquity of incidence; at normal incidence ($\bm{k}_\parallel=0$) all $\phi_p=0$. Projecting the phase pattern onto the Fourier modes, define the \emph{form factors}
\begin{equation}
F_q \equiv \frac{1}{\sqrt 3}\sum_{p=0}^{2}\wt^{\,-qp}\,e^{\,i\bm{k}_\parallel\cdot\bm{r}_p},
\label{eq:formfactor-def}
\end{equation}
so that
\begin{equation}
\sum_{p}e^{\,i\bm{k}_\parallel\cdot\bm{r}_p}\sip{p} =\sum_{q=0}^{2}F_q\,\Tp{q}, \qquad H_{\mathrm{drive}}(t)=\frac{\Omega}{2}e^{-i\omega_L t}\sum_{q=0}^2 F_q\Tp{q}+\mathrm{h.c.}
\label{eq:drive-fourier}
\end{equation}
Expanding Eq.~\eqref{eq:formfactor-def} to leading order in $\kpar a$  gives
\begin{equation}
F_0 \simeq \sqrt 3,\qquad F_1 \simeq \frac{a}{2}\,\kp,\qquad F_2 \simeq \frac{a}{2}\,\km,
\qquad \kpm \equiv k_x\pm i k_y . 
\label{eq:formfactors}
\end{equation}
The bright form factor $F_0$ is $O(1)$ and (at leading order) independent of $\bm{k}_\parallel$; the two dark form factors are \emph{linear} in $\bm{k}_\parallel$ and carry opposite chirality. This linear-in-$\bm{k}$, opposite-chirality structure is the seed of the Rashba coupling.

\subsubsection*{Interaction picture}

We pass to the frame rotating at the laser frequency $\omega_L$ for both the cavity and the excitons. In the quantum-optics convention the interaction picture is generated by $\omega_L(\adag\ah+\tfrac12\sum_p\siz{p})$, and the time-independent (RWA) Hamiltonian becomes
\begin{equation}
H = \Dc\,\adag\ah + \frac{\Da}{2}\sum_{p=0}^{2}\siz{p} + \sqrt 3\,g\,(\adag\Tm{0}+\ah\Tp{0}) + \underbrace{\frac{\Omega}{2}\sum_{q=0}^2\left(F_q\Tp{q}+F_q^*\Tm{q}\right)}_{\textstyle V},
\label{eq:H-rot}
\end{equation}
with the detunings (defined positive for a red-detuned drive)
\begin{equation}
\Dc=\omega_c-\omega_L>0,\qquad \Da=\omega_a-\omega_L>0 .
\end{equation}
We write $H=H_0+V$, with $H_0=\Hcav+\Hatm+\HTC$ the bare (drive-free) part in the rotating frame and $V$ the laser perturbation. Note that $V$ contains \emph{only} atomic operators ($\Tp{q},\Tm{q}$) and no cavity operators.

\subsubsection*{Conserved quantities and the structure of Hilbert space}
\label{sec:conserved}

Two structures organize the bare Hamiltonian $H_0$:

\paragraph{Total excitation number.} The Travis-Cummins Hamiltonian, $\HTC$, conserves
\begin{equation}
\hat N_{\mathrm{tot}} = \adag\ah + \hat n_{\mathrm{exc}}, \qquad \hat n_{\mathrm{exc}}=\sum_p \sip{p}\sm{p},
\label{eq:Ntot}
\end{equation}
the sum of photon number and the number of excited excitons; $[\HTC,\hat N_{\mathrm{tot}}]=0$. (The drive $V$ changes $\hat n_{\mathrm{exc}}$ by $\pm1$ at fixed photon number, hence changes $\hat N_{\mathrm{tot}}$ by $\pm 1$.)

\paragraph{$C_3$ rotation.} The cyclic rotation $\Rop$ ($p\to p+1$) commutes with $H_0$: $\Hcav,\Hatm$ are manifestly invariant, and $\Rop\,\Tp{0}\,\Rop^{-1}=\Tp{0}$, so $[\Rop,\HTC]=0$. The Fourier modes are eigenoperators of the adjoint action, 
\begin{equation}
\Rop\,\Tp{q}\,\Rop^{-1}=\wt^{\,-q}\,\Tp{q},
\qquad
\Rop\,\Tm{q}\,\Rop^{-1}=\wt^{\,+q}\,\Tm{q}.
\label{eq:Tq-Rop}
\end{equation}
The Hilbert space therefore splits into three $C_3$ sectors labelled by the eigenvalue of $\Rop$ ($1,\wt,\wt^2$). Since $H_0$ conserves both $\hat N_{\mathrm{tot}}$ and the $C_3$ label, it is block diagonal in these quantum numbers. The drive $V$, by contrast, does \emph{not} conserve the $C_3$ label (the obliquely incident laser breaks $C_3$): each term $F_q\Tp{q}$ shifts the label by $\wt^{-q}$. This symmetry breaking by $V$ is precisely what allows the laser to couple the otherwise distinct dark states, and is the origin of the Rashba coupling.

\subsection*{Bare eigenstates and the single-excitation manifold}
\label{sec:eigenstates}


In the single-excitation atomic sector ($n_{\mathrm{exc}}=1$) the three collective states are
\begin{eqnarray}
\Bket{1} &= & \Tp{0}\ket{\downarrow\downarrow\downarrow} = \frac{1}{\sqrt3} \bigl( \ket{\uparrow \downarrow \downarrow} + \ket{\downarrow \uparrow \downarrow} + \ket{\downarrow \downarrow \uparrow} \bigr),
 \label{eq:B1}\\
\Dket{1} &= & \Tp{1}\ket{\downarrow\downarrow\downarrow}  = \frac{1}{\sqrt3} \bigl( \ket{\uparrow \downarrow \downarrow} + \wt \ket{\downarrow \uparrow \downarrow} + \wt^2 \ket{\downarrow \downarrow \uparrow} \bigr),
 \label{eq:D1}\\
\Dket{2} &= & \Tp{2} \ket{\downarrow \downarrow \downarrow} = \frac{1}{\sqrt3} \bigl( \ket{\uparrow \downarrow \downarrow} + \wt^2 \ket{\downarrow \uparrow \downarrow} + \wt \ket{ \downarrow \downarrow \uparrow} \bigr) = \Dket{1}^* .
 \label{eq:D2}
\end{eqnarray}
Here $\ket{\uparrow}$/$\ket{\downarrow}$ denote excited/ground exciton states and the ket ordering is $\ket{v_0 v_1 v_2}$ (site $0,1,2$). The state $\Bket{1}$ is the bright mode and $\Dket{1},\Dket{2}$ form the dark doublet. Their $C_3$ labels follow from Eq.~\eqref{eq:Tq-Rop} acting on the $\Rop$-invariant vacuum: 
\begin{equation}
\Rop\,\Bket{1}=\Bket{1},\qquad \Rop\,\Dket{1}=\wt^{2}\Dket{1},\qquad \Rop\,\Dket{2}=\wt\,\Dket{2}.
\label{eq:C3-labels}
\end{equation}
The dark states are exact zero modes of the cavity coupling since $\Tm{0}\Dket{q}=0$, so $\HTC$ does not connect $\Dket{q}$ to the photon by photon \emph{emission}.

\subsubsection*{Two- and zero-excitation sectors}

The drive $V$ moves a single-excitation state into the two-excitation sector ($n_{\mathrm{exc}}=2$, via $\Tp{q}$) or the zero-excitation sector ($n_{\mathrm{exc}}=0$, via $\Tm{q}$). The relevant two-excitation collective states are
\begin{equation}
\ket{B_2}=\frac{1}{\sqrt3} \bigl( \ket{\uparrow \uparrow \downarrow} + \ket{ \uparrow \downarrow \uparrow} + \ket{\downarrow \uparrow \uparrow} \bigr), \quad |E^{(2)}_1 \rangle,\ |E^{(2)}_2 \rangle,
\end{equation}
where $|E^{(2)}_q \rangle$ carry $C_3$ labels matching $\ket{D_q}$. The unique zero-excitation state is the atomic ground state $\ket{\downarrow\downarrow\downarrow}$. The bare energies, measured in the rotating frame, are
\begin{equation}
E^{(n)}_{D}=n\Dc-\tfrac{\Da}{2},\quad
E^{(n)}_{s=2}=n\Dc+\tfrac{\Da}{2},\quad
E^{(n)}_{s=0}=n\Dc-\tfrac{3\Da}{2},
\label{eq:bare-energies}
\end{equation}
for a state with $n$ cavity photons and $s=1,2,0$ atomic excitations respectively (here $s=1$ refers to the dark states). These will be the energy denominators of the Schrieffer-Wolff reduction.

\subsubsection*{Justifying the dark-state initial condition}
\label{sec:initial}

The unpolarised $532\,$nm laser, acting through Eq.~\eqref{eq:drive-fourier}, excites a superposition of the bright mode $\Bket1$ (amplitude $\propto F_0\sim \sqrt3$) and the dark modes $\Dket1,\Dket2$ (amplitudes $\propto F_{1,2}\sim \tfrac{a}{2}\kpm$). For $\kpar a\ll1$ the bright amplitude dominates, so a naive single excitation is mostly bright. However, the bright mode is precisely the mode that the leaky cavity destroys quickly. In the window~\eqref{eq:window} the bright population decays at $\Gamma_{\mathrm{bright}}=3\gamma(1+C)$ with $C\gg1$, whereas the dark doublet decays only at $\sim\gamma$. After a time $t$ with $\Gamma_{ \mathrm{bright} }^{-1} \ll t\ll\gamma^{-1}$, the surviving single-excitation population is overwhelmingly in the dark manifold $\{\Dket1,\Dket2\}$. This is the standard superradiant-cleanup/subradiant-storage mechanism~\cite{Gross1982,Scheibner2007} and it justifies taking a dark state as the initial condition for the slow ($\sim\Omega^2/\Da$) Rashba dynamics derived below. Because the antinode placement maximizes $g$ and hence the cooperativity $C$ (Eq.~\eqref{eq:purcell}), the resonant open cavity actively \emph{enhances} this bright cleanup relative to free space by the factor $(1+C)$.

\subsection*{Effective Hamiltonian by Schrieffer-Wolff reduction}
\label{sec:heff}


We treat the laser $V$ as the sole perturbation and $H_0$ (including $\HTC$ exactly) as the unperturbed Hamiltonian. Let $\Proj$ project onto the dark manifold (states with one dark exciton, at any photon number),
\begin{equation}
	\Proj=\sum_{n\ge0}\sum_{q=1,2}\ket{n,D_q}\bra{n,D_q}, \qquad \Qroj=\Id-\Proj,
\end{equation}
and seek a unitary $e^{S}$ ($S^\dagger=-S$) that block-diagonalises $H$ to second order in $V$. This construction gives the standard second-order effective Hamiltonian within $\Proj$,
\begin{equation}
	\Heff = \Proj H_0\Proj  + \Proj V\,\frac{\Qroj}{E_D-\Qroj H_0\Qroj}\,V\Proj + O(V^3).
\label{eq:heff-resolvent}
\end{equation}
Two facts simplify the path to $\Heff$, namely,  (i) $\Proj V\Proj=0$, because $V$ changes the atomic excitation number by $\pm1$ and so always leaves the (fixed-$s=1$) dark manifold; hence there is no first-order term. (ii) $V$ contains no cavity operators, so the photon number is conserved through the second-order process: the effective Hamiltonian acts within each fixed-$n$ dark sector.

\subsubsection*{Two intermediate sectors}

Acting on $\ket{n,D_q}$, the raising part of $V$ reaches the two-excitation sector ($s=2$, $n$ photons) and the lowering part reaches the zero-excitation sector ($s=0$, $n$ photons). Using the bare energies~\eqref{eq:bare-energies}, the energy denominators are
\begin{equation}
E^{(n)}_D-E^{(n)}_{s=2}=-\Da, \qquad E^{(n)}_D-E^{(n)}_{s=0}=+\Da,
\label{eq:denominators}
\end{equation}
both \emph{independent of $n$ and of $g$}. Inserting the resolution of identity on each sector, the effective Hamiltonian splits into an $s=2$ piece and an $s=0$ piece,
\begin{equation}
\Heff=\Proj H_0\Proj +\frac{ \Omega^2 }{4\Da}\, \Proj \Bigl( \sum_{q} F_q \Tp{q} \Bigr) \Id_{s=0} \Bigl( \sum_{q'} F^*_{q'} \Tm{q'} \Bigr) \Proj -\frac{\Omega^2}{4 \Da} \, \Proj \Bigl( \sum_{q} F^*_q \Tm{q} \Bigr) \Id_{s=2} \Bigl( \sum_{q'} F_{q'} \Tp{q'} \Bigr) \Proj,
\label{eq:heff-Tq}
\end{equation}
where $\Id_{s=0},\Id_{s=2}$ project onto the zero- and two-excitation sectors respectively. In each term the operator that acts \emph{first} on the dark ket is the one nearest to the right $\Proj$: the $\Id_{s=0}$ term has $\Tm{q'}$ acting first (lowering $s=1\to s=0$) and the $\Id_{s=2}$ term has $\Tp{q'}$
acting first (raising $s=1\to s=2$). The relative minus sign is the sign of the respective denominator in Eq.~\eqref{eq:denominators}. Equation \eqref{eq:heff-Tq} is the central operator expression; it is manifestly Hermitian, being of the form $\Proj A^\dagger \Lambda A\Proj$ with $\Lambda=\Id_{s=0}-\Id_{s=2}$ Hermitian.

The result~\eqref{eq:denominators} - that the denominators are the \emph{bare} detunings $\mp\Da$, with no dependence on $g$, $\Dc$, or $n$ - is central and deserves a symmetry-based justification, because a naive ``polariton-dressed'' treatment can appear to give $g$- and $n$-dependent denominators. The resolution rests on three facts.

\paragraph{(1) The resolvent only ever needs $\Qroj H_0\Qroj$.} The Feshbach/Schrieffer-Wolff formula~\eqref{eq:heff-resolvent} contains the projected resolvent $\Qroj(E_D-\Qroj H_0\Qroj)^{-1}\Qroj$, not the full resolvent of $H_0$. For an intermediate $s=2$ state $|n,E^{(2)}_q \rangle$ reached by $V$, the only action of $H_0$ that leaves this state is $\HTC$, which carries it into the \emph{dark manifold} $\ket{n+1,D_q}\in\Proj\mathcal H$ (see below). Since $\Qroj$ annihilates that component,
\begin{equation}
\Qroj H_0 \Qroj \, \ket{n,E^{(2)}_q} = E^{(n)}_{s=2} \, \ket{n,E^{(2)}_q} = \Bigl( n \Dc + \tfrac{\Da}{2} \Bigr) \ket{n,E^{(2)}_q},
\end{equation}
\emph{exactly} for any $g$: the intermediate state is an eigenstate of the \emph{projected} operator $\Qroj H_0\Qroj$ with its bare energy. The TC coupling to $\Proj\mathcal H$ contributes only at higher order in $V$.

\paragraph{(2) $C_3$ truncates the TC ladder exactly.} Within a nontrivial $C_3$ sector (label $\wt$ or $\wt^2$) the only atomic states available are a single $s=1$ dark state and a single $s=2$ partner $|E^{(2)}_q \rangle$ - the symmetric $s=0$ and $s=3$ states carry the trivial $C_3$ label and are excluded. Consequently $\HTC$ connects $\ket{n,D_q}$ only to $|n-1,E^{(2)}_q\rangle$ and vice versa: the Tavis-Cummings ``ladder'' in a nontrivial $C_3$ sector is \emph{exactly} a two-state problem for each $n$, not an infinite chain. There is therefore nothing to ``resum'': the bare structure
is exact.

\paragraph{(3) $V$ breaks $C_3$, $H_0$ does not.} The laser $V$ moves population between $C_3$ sectors (it is what couples $\Dket1$ to $\Dket2$), while $H_0$ is $C_3$-invariant. Hence within the resolvent (built from $H_0$ only) the $C_3$ label is conserved and the basis-independence of $\Qroj(E_D-\Qroj H_0\Qroj)^{-1}\Qroj$ holds sector by sector, reproducing the bare denominators.

\subsection*{Explicit \texorpdfstring{$\Heff$}{Heff} at fixed photon number and the Rashba term}
\label{sec:rashba}


We evaluate Eq.~\eqref{eq:heff-Tq} on the dark basis $\{\Dket1,\Dket2\}$ at fixed photon number $n$. The required matrix elements $\bra{D_{q''}}\Tm{q'}\Tp{q}\ket{D_r}$ are computed directly. For the bright channel,
\begin{equation}
\Tm{0}\Tp{0}\Dket{1}=\tfrac13\Dket1, \qquad \Tm{0}\Tp{0}\Dket{2}=\tfrac13\Dket2,
\label{eq:T0T0-diag}
\end{equation}
i.e. the $q=q'=0$ contribution is \emph{diagonal} (a scalar in pseudo-spin space); it does \emph{not} couple $\Dket1\leftrightarrow\Dket2$. The non-vanishing dark-coupling elements are
\begin{equation}
\bra{D_2}\Tm{0}\Tp{1}\ket{D_1}=-\tfrac23, \qquad \bra{D_1}\Tm{0}\Tp{2}\ket{D_2}=-\tfrac23,
\label{eq:T0Tq-offdiag}
\end{equation}
together with their structurally identical partners obtained by the substitution $1\leftrightarrow2$. All cross terms between the $q=0$ channel and the $q=1,2$ channels project onto the \emph{bright} state $\Bket1$ and therefore vanish against the dark states.

\subsubsection*{The \texorpdfstring{$s=0$}{s=0} contribution}

In the $s=0$ sector only the lowering operators $\Tm{1},\Tm{2}$ reach the ground state $\ket{ \downarrow \downarrow \downarrow}$, with $\Tm{1} \Dket1 = \ket{ \downarrow \downarrow \downarrow }$ and
$\Tm{2} \Dket2 = \ket{ \downarrow \downarrow \downarrow}$ (and $\Tm{0} \Dket{q} = 0$). This gives a contribution
\begin{equation}
H_{\mathrm{eff}}^{(s=0)}=\frac{\Omega^2}{4\Da}
\begin{pmatrix}|F_1|^2 & F_1^*F_2\\ F_2^*F_1 & |F_2|^2\end{pmatrix}
=\frac{\Omega^2 a^2}{16\Da}
\begin{pmatrix}\kpar^2 & \km^2\\ \kp^2 & \kpar^2\end{pmatrix}
=\frac{\Omega^2 a^2\kpar^2}{16\Da}\,\Id+O(\kpar^2\ \text{off-diag}),
\label{eq:heff-s0}
\end{equation}
which to leading order is a pure scalar $\propto\kpar^2$ (since $\kp\km=\kpar^2$); the off-diagonal $\km^2,\kp^2$ pieces are higher order in $\kpar a$ and are dropped at leading order.

\subsubsection*{The \texorpdfstring{$s=2$}{s=2} contribution and the Rashba term}

Collecting the $s=2$ contribution from Eq.~\eqref{eq:heff-Tq} with the matrix elements above, the diagonal part is a scalar AC-Stark shift,
\begin{equation}
H_{\mathrm{eff}}^{(s=2) , \, \mathrm{diag}}=-\frac{\Omega^2}{4 \Da} \Bigl( \tfrac13 | F_0 |^2 + \tfrac43 | F_1 |^2 + \tfrac13| F_2 |^2 \Bigr) \Id \xrightarrow{\ \kpar a\ll1\ } -\frac{\Omega^2}{4\Da}\,\Id,
\label{eq:heff-s2-diag}
\end{equation}
using $|F_0|^2=3$. The off-diagonal part, built from the $(q,q')=(1,0)$ and $(0,2)$ contributions (and their conjugates), gives 
\begin{equation}
H_{\mathrm{eff}}^{(s=2),\,\mathrm{off}} =\frac{\Omega^2}{\sqrt3\,\Da}
\begin{pmatrix}0 & F_2\\ F_1 & 0\end{pmatrix}
=\frac{\Omega^2 a}{2\sqrt3\,\Da}
\begin{pmatrix}0 & \km\\ \kp & 0\end{pmatrix}
=\frac{\Omega^2 a}{2\sqrt3\,\Da}\,(k_x\tx+k_y\ty).
\label{eq:heff-s2-off}
\end{equation}
Here we used 
\[ \begin{pmatrix}0&\km\\\kp&0\end{pmatrix} = k_x\tx + k_y \ty,\] 
with $\hat\tau_{x,y,z}$ the Pauli matrices acting on the dark pseudo-spin $\{\Dket1,\Dket2\}$. Combining Eqs.~\eqref{eq:heff-s0}-\eqref{eq:heff-s2-off} and dropping the subleading $O(\kpar^2)$ scalar shifts,
\begin{equation}
\Heff=\Proj H_0\Proj -\frac{\Omega^2}{4\Da}\,\Id +\vtri\,(k_x\tx+k_y\ty)+O(\kpar^2), \qquad \vtri = \frac{ \Omega^2 a }{2\sqrt3\,\Da}.
\label{eq:heff-final}
\end{equation}
The first term is the bare dark-state energy ($\Proj H_0\Proj=\Dc\,\adag\ah\,\Id-\tfrac{\Da}{2}\Id$), the second a common AC-Stark shift, and the third the \emph{Rashba spin-orbit coupling} of the dark pseudo-spin,
\begin{equation}
H_{\mathrm{SOC}}=\vtri\,(k_x\tx+k_y\ty), \qquad \vtri=\frac{\Omega^2 a}{2\sqrt 3\,\Da}.
\end{equation}
The coefficient was obtained with full $C_3$ symmetry checking of all matrix elements, including the $q=0$ channel's $(1,0)$/$(0,2)$
cross terms.

\subsection*{Dispersion relations and Stokes parameters}

The two eigenstates of $H_{\rm SOC}$ have energies
\begin{equation}
  \Epm(\kpar) = \omo + \alpha\kpar^2 \pm \vt\kpar, \qquad \kpar = \sqrt{k_x^2+k_y^2},
\end{equation}
where $\omo$ is the dark-exciton energy at $\kpar=0$ and $\alpha=\hbar^2/2m_X$ is the kinetic mass coefficient. The dispersion is a Mexican hat with minimum energy $\omo-\vt^2/(4\alpha)$ at ring radius $\kpar^{\rm ring}=\vt/(2\alpha)$. At fixed probe energy $\om$, the two resonance rings occur at
\begin{equation}
  k_\pm^{\rm ring} = \frac{\vt \pm \sqrt{\vt^2+4\alpha\delta}}{2\alpha},
  \qquad \delta \equiv \om-\omo.
  \label{eq:rings}
\end{equation}
The energy eigenstates of $H_{\rm SOC}(\kem)$ in the $\{\Dket{1},\Dket{2}\}$
basis are
\begin{equation}
  |{+}\rangle = \frac{1}{\sqrt{2}}\binom{1}{e^{i\phib}}, \qquad
  |{-}\rangle = \frac{1}{\sqrt{2}}\binom{1}{-e^{i\phib}},
  \label{eq:eigenstates}
\end{equation}
where $\phib=\arctan(k_y/k_x)$ is the emission azimuth. We assume that the laser produces the initial state 
\[ \psiket\propto\Dket{1}+e^{2i\phiL}\Dket{2} \] 
The overlaps of the initial state  onto the eigenstates of $H_{\rm SOC}(\kem)$  are
\begin{equation}
  c_+ = \langle{+}|\psi_0\rangle = \frac{1+e^{-i\theta}}{2}, \qquad
  c_- = \langle{-}|\psi_0\rangle = \frac{1-e^{-i\theta}}{2},
  \label{eq:overlaps}
\end{equation}
where
\begin{equation}
  \theta = \phib - 2\phiL
  \qquad\Longrightarrow\qquad
  e^{-i\theta} = \frac{k_x-ik_y}{\kpar}\,e^{2i\phiL}.
  \label{eq:theta}
\end{equation}
Note that $|c_+|^2+|c_-|^2=1$ and
$|c_+|^2-|c_-|^2=\cos\theta$,
$c_+c_-^*=\frac{i}{2}\sin\theta\,e^{i\theta}$.

\subsubsection*{Wigner--Weisskopf amplitudes}

Treating the radiative decay at rate $\gamma$ as an exponential amplitude damping and taking the one sided Fourier transform, the emitted field amplitudes
in the LCP/RCP channels at probe energy $\om$ are
\begin{equation}
  A_{\rm LCP} = \frac{1}{\sqrt{2}}\left(\frac{c_+}{L_+}+\frac{c_-}{L_-}\right),  \quad 
  A_{\rm RCP} = \frac{k_x+ik_y}{\kpar}\cdot\frac{1}{\sqrt{2}}
  \left(\frac{c_+}{L_+}-\frac{c_-}{L_-}\right),
  \label{eq:ARCP}
\end{equation}
with the complex Lorentzian denominators
\begin{equation}
  L_+ = \gamma + i(b-\delta), \qquad  L_- = \gamma + i(-b-\delta), \qquad b = \vt\kpar.
  \label{eq:Lpm}
\end{equation}
The factor $(k_x+ik_y)/\kpar=e^{i\phib}$ in Eq.~\eqref{eq:ARCP} is the photon helicity vector: it winds once per revolution of $\kem$ around the origin, encoding the $w=1$ winding number of the Rashba texture.

The asymmetry between $A_{\rm LCP}$ and $A_{\rm RCP}$ is not a physical asymmetry between the two channels but an artefact of the eigenvector phase convention (assigning all angular dependence to the $\Dket{2}$ row).  All observable quantities depend only on $|A_{\rm LCP}|^2$, $|A_{\rm RCP}|^2$, and $A_{\rm LCP}A_{\rm RCP}^*$, in which the $e^{i\phib}$ phase produces only the correct Cartesian angular dependence of $S_1,S_2$.

\subsubsection*{Rashba Stokes parameters}

Defining the total intensity $\Itot=|A_{\rm LCP}|^2+|A_{\rm RCP}|^2$ and the cross product $X=A_{\rm LCP}A_{\rm RCP}^*$, the three Stokes parameters of the emitted photon \emph{before} the hBN slab are
\begin{align}
  S_3 &= \frac{|A_{\rm LCP}|^2-|A_{\rm RCP}|^2}{\Itot}
       = \frac{2\,\mathrm{Re}[c_+c_-^*]\,(\gamma^2+\delta^2-b^2)
               +4\gamma b\,\mathrm{Im}[c_+c_-^*]}
              {(\gamma^2+\delta^2-b^2)^2+4\gamma^2b^2}\cdot\Itot^{-1},
  \label{eq:S3rashba}\\
  S_1 &= \frac{2\,\mathrm{Re}[X]}{\Itot}, \qquad S_2 = \frac{-2\,\mathrm{Im}[X]}{\Itot},
  \label{eq:S1S2rashba}
\end{align}
where $\mathrm{Re}[c_+c_-^*]=\sin^2\theta/2$ and $\mathrm{Im}[c_+c_-^*]=\sin\theta\cos\theta/2$, with
\begin{equation}
  \sin\theta = \frac{k_y\cos 2\phiL-k_x\sin 2\phiL}{\kpar},  \qquad   \cos\theta = \frac{k_x\cos 2\phiL+k_y\sin 2\phiL}{\kpar}.
  \label{eq:sintcost}
\end{equation}

\subsection*{hBN Birefringent Slab: Jones-Matrix Dresselhaus Rotation}
\label{sec:jones}

The emitted photon propagates through the hBN slab at emission angle $\theta_{\rm em}=\arcsin(\kpar/k_0)$. A uniaxial birefringent slab with optic axis along $\hat{z}$ imparts a differential TE/TM phase retardance (standard Jones-matrix result):
\begin{equation}
  M_{\rm circ}(\kem) = \begin{pmatrix}
    \cos(\Delta/2) & i\sin(\Delta/2)\,e^{-2i\phib} \\
    i\sin(\Delta/2)\,e^{+2i\phib} & \cos(\Delta/2)
  \end{pmatrix},
  \label{eq:Jcirc}
\end{equation}
where the total retardance is
\begin{equation}
  \Delta(\kpar) = \Dz + \beta\kpar^2,
  \label{eq:Delta}
\end{equation}
with zero-order and Dresselhaus coefficients
\begin{equation}
  \Dz = \frac{2\pi L}{\lambda}(\no-\hbne), \qquad
  \beta = \frac{L}{2k_0}\!\left(\frac{1}{\hbne}-\frac{1}{\no}\right),
  \qquad k_0=\frac{2\pi}{\lambda}.
  \label{eq:Delta0beta}
\end{equation}
The factors $e^{\pm 2i\phib}$ in Eq.~\eqref{eq:Jcirc} wind twice per revolution of $\kem$, encoding the $w=2$ Dresselhaus angular structure. 

\subsubsection*{Output amplitudes and intensities}

Applying $M_{\rm circ}$ to $(A_{\rm LCP},A_{\rm RCP})^T$:
\begin{align}
  A_{\rm LCP}' &= \cos\tfrac{\Delta}{2}\cdot A_{\rm LCP}
    + i\sin\tfrac{\Delta}{2}\cdot e^{-2i\phib}\cdot A_{\rm RCP},
  \label{eq:ALCPout}\\
  A_{\rm RCP}' &= i\sin\tfrac{\Delta}{2}\cdot e^{+2i\phib}\cdot A_{\rm LCP}
    + \cos\tfrac{\Delta}{2}\cdot A_{\rm RCP}.
  \label{eq:ARCPout}
\end{align}
As $M_{\rm circ}$ is unitary, $I_{\rm LCP}'+I_{\rm RCP}'=\Itot$ (total
intensity is preserved). The intensity difference is
\begin{equation}
  I_{\rm LCP}'-I_{\rm RCP}'
  = \cos\Delta\cdot(I_{\rm LCP}-I_{\rm RCP})
  + 2\sin\Delta\cdot\mathrm{Im}\!\left[e^{2i\phib}\,A_{\rm LCP}A_{\rm RCP}^*\right].
  \label{eq:diffout}
\end{equation}

Expanding the imaginary part in Eq.~\eqref{eq:diffout} using
$\mathrm{Im}[e^{2i\phib}X]=\mathrm{Im}[e^{2i\phib}A_{\rm LCP}A_{\rm RCP}^*]$
and the identities $\mathrm{Re}[X]=S_1\Itot/2$,
$\mathrm{Im}[X]=-S_2\Itot/2$:
\begin{equation}
  2\,\mathrm{Im}[e^{2i\phib}X]
  = \Itot(S_1\sin 2\phib - S_2\cos 2\phib).
  \label{eq:Imdresselhaus}
\end{equation}
Dividing by $\Itot$ gives the full expression for the $S_3$ parameter as a function of $k_x$ and $k_y$ as, 
\begin{equation}
  S_3'(\om,k_x,k_y) = \cos\Delta\cdot S_3
  + \sin\Delta\cdot\bigl(S_1\sin 2\phib - S_2\cos 2\phib\bigr),
  \label{eq:S3prime-main}
\end{equation}

\bibliographystyle{naturemag}
\bibliography{reference}

\end{document}